\documentstyle[12pt]{article}

\newcommand{\newsection}{\setcounter{equation}{0}\section}

\newcommand{\tr}{\mbox{tr}}
\newcommand{\Tr}{\mbox{Tr}}

\newcommand{\complex}{{\bf C}} 
\newcommand{\zed}{{\bf Z}} 
\newcommand{\real}{{\bf R}} 
\newcommand{\SK}{{\bb K}} 
\newcommand{\SA}{{\bb A}} 
\newcommand{\sSA}{{\bbs A}} 
\newcommand{\SS}{{\bb S}} 
\newcommand{\sSS}{{\bbs S}} 
\newcommand{\SW}{{\bb W}} 
\newcommand{\SE}{{\bb E}} 
\newcommand{\sSE}{{\bbs E}} 
\newcommand{\SD}{{\bb D}} 
\newcommand{\SlashA}{{\cal A}\!\!\!\!/\,}
\newcommand{\SlashN}{\nabla\!\!\!\!/\,}
\newcommand{\Dirac}{{\cal D}\!\!\!\!/\,} 
\newcommand{\Sdirac}{\SD\!\!\!\!/\,} 
\newcommand{\dirac}{{D\!\!\!\!/\,}} 
\newcommand{\id}{{1\!\!1}} 
\newcommand{\orbit}{{\bb M}} 
\newcommand{\cliff}{C\!\ell} 
\newcommand{\ind}{{\rm index}} 
\newcommand{\Ind}{{\rm Ind}} 
\newcommand{\Index}{{\rm Index}} 
\newcommand{\ch}{{\rm ch}} 
\newcommand{\klein}{{(-1)^{F_{\rm L}}}} 

\newcommand{\K}{{\rm K}} 
\newcommand{\KK}{{\rm KK}} 
\newcommand{\hil}{{\cal H}} 
\newcommand{\alg}{{\cal A}} 
\newcommand{\fred}{{\cal F}} 
\newcommand{\fredadj}{{\cal Q}} 
\newcommand{\comp}{{\cal K}(\hil)} 
\newcommand{\extra}{{{\bf S}^1}} 
\newcommand{\toe}{{\cal T}} 

\font\mybb=msbm10 at 12pt
\def\bb#1{\hbox{\mybb#1}}
\font\mybbs=msbm10 at 9pt
\def\bbs#1{\hbox{\mybbs#1}}

\def\e{{\rm e}}
\def\beq{\begin{equation}}
\def\eeq{\end{equation}}
\def\bea{\begin{eqnarray}}
\def\eea{\end{eqnarray}}
\newcommand{\nn}{\nonumber}
\def\bd{\begin{displaymath}}
\def\ed{\end{displaymath}}

\makeatletter
\newdimen\normalarrayskip              
\newdimen\minarrayskip                 
\normalarrayskip\baselineskip
\minarrayskip\jot
\newif\ifold             \oldtrue            \def\new{\oldfalse}
\def\arraymode{\ifold\relax\else\displaystyle\fi} 
\def\@arrayskip{\ifold\baselineskip\z@\lineskip\z@
     \else
     \baselineskip\minarrayskip\lineskip2\minarrayskip\fi}
\def\@arrayclassz{\ifcase \@lastchclass \@acolampacol \or
\@ampacol \or \or \or \@addamp \or
   \@acolampacol \or \@firstampfalse \@acol \fi
\edef\@preamble{\@preamble
  \ifcase \@chnum
     \hfil$\relax\arraymode\@sharp$\hfil
     \or $\relax\arraymode\@sharp$\hfil
     \or \hfil$\relax\arraymode\@sharp$\fi}}
\def\@array[#1]#2{\setbox\@arstrutbox=\hbox{\vrule
     height\arraystretch \ht\strutbox
     depth\arraystretch \dp\strutbox
     width\z@}\@mkpream{#2}\edef\@preamble{\halign \noexpand\@halignto
\bgroup \tabskip\z@ \@arstrut \@preamble \tabskip\z@ \cr}%
\let\@startpbox\@@startpbox \let\@endpbox\@@endpbox
  \if #1t\vtop \else \if#1b\vbox \else \vcenter \fi\fi
  \bgroup \let\par\relax
  \let\@sharp##\let\protect\relax
  \@arrayskip\@preamble}
\makeatother

\begin{document}

\renewcommand{\footnotesize}{\small}

\thispagestyle{empty}

\begin{flushright}
\baselineskip=12pt
HWM--01--31\\
NBI--HE--00--07\\
EMPG--01--11\\
hep-th/0108043\\
\hfill{  }\\ August 2001\\
Revised October 2001
\end{flushright}

\vskip 0.5cm

\begin{center}

\baselineskip=24pt

{\Large\bf Superconnections, Anomalies\\ and Non-BPS Brane Charges}\\[15mm]

\baselineskip=12pt

{\bf Richard J. Szabo}\\[5mm]

{\it Department of Mathematics, Heriot-Watt University \footnote{Permanent
Address.}\\ Riccarton, Edinburgh EH14 4AS, Scotland}\\[2mm] and \\[2mm] {\it
The Niels Bohr Institute\\ Blegdamsvej 17, DK-2100 Copenhagen \O,
Denmark}\\[2mm] {\tt R.J.Szabo@ma.hw.ac.uk}\\[30mm]

{\sc Abstract}

\begin{center}
\begin{minipage}{14cm}

\baselineskip=12pt

The properties of brane-antibrane systems and systems of unstable D-branes in
Type II superstring theory are investigated using the formalism of
superconnections. The low-energy open string dynamics is shown to be probed by
generalized Dirac operators. The corresponding index theorems are used to
compute the chiral gauge anomalies in these systems, and hence their
gravitational and Ramond-Ramond couplings. A spectral action for the
generalized Dirac operators is also computed and shown to exhibit precisely the
expected processes of tachyon condensation on the brane worldvolumes. The
Chern-Simons couplings are thereby shown to be naturally related to Fredholm
modules and bivariant K-theory, confirming the expectations that D-brane charge
is properly classified by K-homology.

\end{minipage}
\end{center}

\end{center}

\vfill
\newpage

\tableofcontents

\vfill
\newpage

\setcounter{page}{1}

\newsection{Introduction and Summary}

The surge of interest in recent years in the study of unstable systems of
D-branes (see~\cite{NonBPSrev} for reviews) has been sparked in part because
they carry information about the non-perturbative vacuum of string theory. The
instability present in such systems is marked by the appearence of tachyonic
modes in the open string sectors. It was originally suggested that the tachyon
field on these D-branes should be properly understood as a Higgs
field~\cite{Sen1}, and that the instability is due to a perturbative expansion
about an unstable extremum of the tachyon potential. Assuming certain
properties of the potential, the system may then decay into its true vaccum
state and sometimes leave behind a topological defect which is
indistinguishable from a stable, supersymmetric D-brane. The assumptions which
go into this argument have been subsequently verified within the framework of
string field theory~\cite{3SFT}--\cite{TTU} and of worldsheet $\sigma$-model
effective actions~\cite{sigmaeff}.

The prototypical system consists of a collection of branes and antibranes. It
has been used to develop the topological classification of D-brane charges in
terms of the cohomology of their Chan-Paton gauge bundles, namely
K-theory~\cite{MMK,WittenK} (see~\cite{osrev} for a review). In this setting,
Ramond-Ramond (RR) charge is characterized by the formal difference of the
vector bundles carried by the branes and antibranes. There is a canonical map
from the K-theory group into cohomology given by the Chern character, which can
be used to give explicit charge formulas for the coupling of Ramond-Ramond
potentials to the D-brane worldvolumes. For this, the gauge connections should
be replaced by the appropriate analogs for virtual bundles. It was suggested
in~\cite{WittenK} that the appropriate geometric extension is given by a
superconnection~\cite{Quillen1}--\cite{BGV} which naturally incorporates the
tachyon field of the unstable system. This description has been exploited
recently to describe many aspects of the couplings of unstable D-branes to
closed string supergravity fields~\cite{KLDbarD,TTU},\cite{kw}--\cite{AGS}.

In this paper we will present a detailed, mathematical exposition of these
relationships. We will pay particular attention to the systematic derivation of
the couplings of the unstable D-branes to the Ramond-Ramond tensor potentials.
Our analysis will rely on the identifications of anomalies in the brane
worldvolume quantum field theories due to the presence of chiral fermion
fields. The RR charges of these systems may then be determined by the
appropriate modification of standard index theoretical techniques. Such
interpretations have also been pointed out in~\cite{KLDbarD,SchwarzW}.

Recall that the RR couplings on the worldvolume $\Sigma$ of $N$ coincident
D-branes in Type~II superstring theory is expressed through the Wess-Zumino
action~\cite{MMK},\cite{RRD}--\cite{ss}
\beq
S_{\rm WZ}=\int\limits_\Sigma{\cal C}
\wedge\tr^{~}_N~\e^{(F_A-{\cal B})/2\pi i}\wedge
\e^{d/2}\wedge\sqrt{\frac{\widehat{A}(R_T)}{\widehat{A}(R_N)}} \ ,
\label{SWZDbrane}\eeq
where throughout we shall work in appropriate string units. In
(\ref{SWZDbrane}), $\cal C$ is the pullback of the total RR form potential $C$
under the worldvolume embedding $\phi:\Sigma\to X$ into the spacetime manifold
$X$, $F_A$ is the field strength of the $U(N)$ gauge field $A$ on the branes,
with $\tr^{~}_N$ the trace in the fundamental representation of $U(N)$, and
$\cal B$ is the pullback of the NS-NS two-form potential $B$. The quantities
$R_T$ and $R_N$ are the curvatures of the tangent and normal bundles to
$\Sigma$ in $X$, respectively, $\widehat{A}(R)$ is the usual Dirac index, and
$d$ is a degree two characteristic class which defines a spin$^c$ structure on
$\Sigma$. This requires the NS-NS $B$-field to be topologically
trivial~\cite{WittenK,WittenAdS,FW}. The action (\ref{SWZDbrane}) can be
written in terms of spacetime quantities alone in the form~\cite{MMK}
\beq
S_{\rm WZ}=\int\limits_XC\wedge\ch(\phi^{~}_!E)\wedge
\sqrt{\widehat{A}(TX)} \ ,
\label{SWZKtheory}\eeq
where $\phi^{~}_!$ is the induced Gysin map acting on K-theory, $E$ is the
Chan-Paton gauge bundle supported by the D-branes, and ch denotes the usual
Chern character. This leads to an interpretation of RR-charges as elements
$x=\phi^{~}_!E\in\K^0(X)$ of the K-theory of spacetime. A similar property is
true of the Ramond-Ramond fields themselves~\cite{MooreWitten,FreedHop}. For
non-abelian gauge bundles, the couplings (\ref{SWZDbrane}) and
(\ref{SWZKtheory}) require appropriate modifications in order that the action
be T-duality invariant~\cite{myers,hassan}.

For a system of $N^+$ coincident branes and $N^-$ coincident antibranes
wrapping a submanifold $\Sigma$ of spacetime, we will derive the charge formula
\bea
S_{\rm WZ}^{D\overline{D}}&=&\int\limits_\Sigma{\cal C}\wedge
\tr^{~}_{N^+\oplus N^-}\pmatrix{\id_{N^+}&0\cr0&-\id_{N^-}\cr}
\exp\frac1{2\pi i}\,
\pmatrix{F_{A^+}+T^\dagger T&(DT)^\dagger\cr DT&F_{A^-}+TT^\dagger\cr}
\nn\\&&\wedge\,\e^{-{\cal B}/2\pi i}\wedge\e^{d/2}\wedge
\sqrt{\frac{\widehat{A}(R_T)}{\widehat{A}(R_N)}} \ ,
\label{SWZDbarD}\eea
where $F_{A^\pm}$ are the field strengths of the $U(N^\pm)$ gauge fields
$A^\pm$ on the branes and antibranes, respectively, $T$ is the bi-fundamental
$\overline{\bf N^-}\otimes{\bf N^+}$ tachyon field, and the gauge covariant
derivative is given by $DT=dT+A^-T+TA^+$. The gauge field part of this action
in the case $N^+=N^-$ was obtained to leading orders in powers of the tachyon
field in~\cite{kw} by a tree-level calculation of the open string effective
action on the brane-antibrane system. As proposed there, the full gauge
coupling can be elegantly expressed in terms of a superconnection associated
with the open string fields $A^\pm$ and $T$, and indeed within the present
formalism this is how these terms shall emerge. For the appropriate Higgs
profile of the tachyon field, these couplings were shown to yield the
anticipated RR couplings to lower dimensional D-branes that remain after
tachyon condensation. We will show that the same is true when one includes the
gravitational couplings to the brane worldvolume, as in (\ref{SWZDbarD}). The
action (\ref{SWZDbarD}) also has an appropriate modification which makes it
explicitly T-duality invariant~\cite{TTU}.

There is an analogous story for the unstable D$p$-branes of Type II superstring
theory which occur at the ``wrong'' values of $p$~\cite{Sen4,Horava}. From the
form of the corresponding open string scattering amplitudes, it was proposed
that a system of $N$ unstable branes wrapping a submanifold $\Sigma$ of
spacetime generates a Ramond-Ramond coupling of the
form~\cite{Sen4}--\cite{bcr}
\beq
\tilde S_{\rm WZ}^{(0)\rm non}=\int\limits_\Sigma{\cal C}\wedge d\,
\tr^{~}_N\,T~\e^{(F_A-{\cal B})/2\pi i}\wedge
\e^{d/2}\wedge\sqrt{\frac{\widehat{A}(R_T)}{\widehat{A}(R_N)}} \ ,
\label{SWZnon0}\eeq
where here $T$ is the Hermitian tachyon field which lives on the non-BPS branes
and which belongs to the adjoint representation of the $U(N)$ gauge group. This
form of the RR coupling has been shown to correctly reproduce, modulo the
gravitational couplings, the charge formula for BPS D-branes after tachyon
condensation. However, as in (\ref{SWZDbarD}), one expects interactions between
the gauge fields and higher powers of the tachyon field~\cite{kw}. Based on
this observation, it was proposed that the full Wess-Zumino action describing
this coupling is given by~\cite{kluson1}
\beq
\tilde S_{\rm WZ}^{\rm non}=\int\limits_\Sigma{\cal C}\wedge\sum_{n,m\geq0}
a_{nm}~\tr^{~}_N\left(D_AT\right)^{2n+1}\,T^{2m}\wedge
\e^{(F_A-{\cal B})/2\pi i}
\wedge\e^{d/2}\wedge\sqrt{\frac{\widehat{A}(R_T)}{\widehat{A}(R_N)}} \ ,
\label{SWZnon}\eeq
where $a_{nm}$ are undetermined numerical coefficients. The gauge covariant
derivative in (\ref{SWZnon}) is given by $D_AT=dT+[A,T]$. Again, modulo the
gravitational parts, the action (\ref{SWZnon}) correctly reproduces the
Wess-Zumino term for BPS D-branes after tachyon condensation. It likewise
admits a T-duality invariant extension~\cite{TTU,myersnon}.

In what follows we shall show that this is indeed the case. We will find that
the explicit realization of the proposed expansion (\ref{SWZnon}) is given by
an RR-coupling of the form
\bea
S_{\rm WZ}^{\rm non}&=&\int\limits_\Sigma{\cal C}\wedge\tr^{~}_N\left\{
\exp\left(\frac1{2\pi i}\,\Bigl(T^2+D_AT\Bigr)+\sum_{r,s=1}^\infty
\Xi_{rs}[A,T]\right)\right.\nn\\& &-\left.\exp\left(\frac1{2\pi i}\,
\Bigl(T^2-D_AT\Bigr)+\sum_{r,s=1}^\infty\Xi_{rs}[A,-T]\right)\right\}\wedge
\e^{(F_A-{\cal B})/2\pi i}\wedge\e^{d/2}\wedge
\sqrt{\frac{\widehat{A}(R_T)}{\widehat{A}(R_N)}} \ , \nn\\&&
\label{SWZnonfull}\eea
where $\Xi_{rs}[A,T]$ are involved invariant functions of the gauge and tachyon
fields on the system of non-BPS D-branes. They are given explicitly in
(\ref{calCrsdef}) below. In fact, we will find that the lowest order couplings
(\ref{SWZnon0}) are induced by regarding the system of unstable D-branes as the
naive dimensional reduction of a stable system in one higher dimension.
Instead, the full action (\ref{SWZnonfull}) comes from realizing the unstable
branes in terms of a certain projection of a brane-antibrane
system~\cite{Sen4,BHYi}, and hence from the superconnection formalism that is
appropriate to such systems~\cite{KLDbarD,TTU,AIO}. In this sense, the formula
(\ref{SWZnon0}) may be thought of as a sort of ``field theoretical'' limit of
the ``stringy'' action (\ref{SWZnonfull}).

While the boundary string field formalism yields explicit forms for the
Ramond-Ramond couplings of unstable systems of D-branes that give the correct
brane tensions in processes involving tachyon condensation~\cite{KLDbarD,TTU},
we will take a more mathematical approach to the construction of the effective
actions for these systems. The present approach focuses more closely on a set
of generalized Dirac operators associated to the unstable D-branes, which yield
an equivalent description of their geometry as that by superconnections and
which serve as a probe of the low-energy open string dynamics. With this
analysis, we will develop an intuitive, geometric understanding of the role of
the tachyon field on non-BPS D-branes, and hence to the origins of their
worldvolume effective field theories. A real virtue of the present formalism is
that it enables the construction of {\it global} expressions for the
worldvolume actions. In addition to yielding mathematical constructions of the
Ramond-Ramond couplings, the Dirac operators yield very compact, spectral forms
for the kinetic parts of the effective action describing the propagation of the
worldvolume fields. Such generalized Dirac-Born-Infeld type actions have been
previously proposed in~\cite{AIO,Sen5,kluson2}. The conditions required for
tachyon condensation are thereby reproduced in a very natural way. We will show
that the RR couplings for such generalized vortex configurations reduce to the
anticipated forms for supersymmetric D-branes, generalizing earlier
calculations to incorporate the worldvolume gravitational couplings. We will
also describe the modifications of these actions required for T-duality
invariance, although at present we do not have a natural geometric origin for
these extra couplings.

Although for most of our analysis we restrict attention to D-branes in Type II
superstring theory for simplicity, the techniques developed can be generalized
to other string theories and to other brane systems (See~\cite{SchwarzW} for
further examples). In particular, a system of stable, supersymmetric D-branes
is also a special instance of this general formalism. The formalism of
generalized Dirac operators then suggests another intriguing relationship
between Ramond-Ramond charges and K-theory. More precisely, it leads
immediately to an interpretation of D-brane charge in terms of bivariant
K-theory, and hence analytic K-homology. The relationship unveiled here
complements previous considerations which suggest that D-brane charge should
really be associated to K-homology~\cite{KDhomology,NCtachyon}. As we will
show, the formalism of analytic K-homology takes a particularly transparent
form when applied to systems of unstable D-branes.

The structure of the remainder of this paper is as follows. In the next section
we will give a mathematical introduction to the theory of superconnections and
how they naturally describe the geometry of brane-antibrane systems. We also
introduce the generalized Dirac operators which will play a prominent role in
this paper, and also how the standard formalism for dealing with gauge
anomalies can be extended to this case. In section~3 we derive the form of the
Ramond-Ramond couplings on brane-antibrane systems by using index theoretical
arguments based on the identification of gauge anomalies generalized to the
case of superconnection gauge fields. We describe the modifications of these
actions due to non-abelian structure groups and topologically trivial
$B$-fields. We also describe how tachyon condensation processes emerge
naturally within this geometrical formalism, and how the standard anomalous
couplings on systems of stable D-branes are reproduced from these RR-couplings
via the global bound state construction. In section~4 we then turn to the
derivation of the Ramond-Ramond couplings on systems of unstable D-branes. We
show how the leading order terms in the tachyon field arise from a dimensional
reduction of a stable system in one higher dimension. We then relate the
couplings to the formalism of superconnections by deriving the full expansion
in powers of the tachyon field via reduction from a brane-antibrane system. The
resulting actions are also shown to reduce appropriately for Higgs profiles of
the tachyon field. In section~5 we indicate how the couplings universally
extend to all branes of Type II superstring theory and M-theory, and hence
illustrate the power of the analysis in that it can be applied to many other
string theories and brane systems than the ones considered in this paper.
Finally, in section~6 we investigate the intimate relationship between D-brane
charges and K-theory in light of the present analysis. We show that the
formalism of superconnections and generalized Dirac operators leads naturally
to the interpretation of D-brane charges in terms of Fredholm modules and
bivariant K-theory, and hence analytic K-homology. We also show how the
reductions leading to the charge formulas for unstable D-branes admit natural
interpretations within this homological framework, thereby lending further
support to the suggestion that D-brane charge should be properly understood
within the context of K-homology.

\newsection{Superconnections on Brane-Antibrane Systems}

In this section we will describe the geometry of brane-antibrane systems from
mostly a mathematical perspective, primarily to introduce notions that will
play a fundamental role in later sections. Such systems are $\zed_2$-graded
objects \cite{WittenK} and are thereby most naturally described using the
language of superconnections \cite{Quillen1,MQ} (See \cite{BGV} for a concise
introduction). Superconnections were originally introduced as geometric objects
associated with graded vector bundles whereby the conventional integer grading
by differential form degree is replaced with a $\zed_2$-grading, giving more
freedom to the standard constructions of differential geometry. In
\cite{Quillen1} the Chern-Weil invariants of a superbundle were constructed and
the definition of the Chern character of a superconnection was given. We will
focus primarily on the basic result that a certain class of superconnections
are in a one-to-one correspondence with (generalized) Dirac operators, a fact
that will be at the heart of the analysis of this paper. Earlier descriptions
of the geometry of Higgs fields using the formalism of superconnections can be
found in \cite{Neeman}.

\subsection{Chan-Paton Superbundles}

Consider a system of $N^+$ coincident D$p$-branes and $N^-$
D$\overline{p}$-branes in Type II superstring
theory.\footnote{\baselineskip=12pt For $p=9$, tadpole anomaly cancellation
requires there to be an equal number of spacetime filling branes and antibranes
in the Type IIB vacuum state and $N^+=N^-$.} We assume that the branes all wrap
a common worldvolume $\Sigma$ of dimension $p+1$ in the ten-dimensional
spacetime manifold $X$ which is endowed with a spin structure and a Riemannian
metric. We are interested in the properties of Chan-Paton gauge bundles
$E\to\Sigma$ over the brane-antibrane worldvolume. The open string Hilbert
space of the brane-antibrane system has a natural $\zed_2$-grading which may be
associated to it \cite{WittenK}. The Chan-Paton gauge group of the
$p$-$\overline{p}$ pairs is $U(N^++N^-)$ and it acts on the Hilbert space
${\cal H}=\complex^{N^++N^-}$. It has an index $a=+$ for an open string ending
on a $p$-brane and $a=-$ for an open string ending on a $\overline{p}$-brane.
The endpoints of the $p$-$\overline{p}$ open strings therefore carry a charge
which takes values in a graded quantum Hilbert space ${\cal H}={\cal
H}^+\oplus{\cal H}^-$. We may regard the $a=+$ state as bosonic and $a=-$ as
fermionic.

This implies that any complex vector bundle $E\to\Sigma$ inherits this
$\zed_2$-grading and becomes a superbundle $E=E^+\oplus E^-$, i.e. a bundle
whose fibers $E_x=E_x^+\oplus E_x^-$, $x\in\Sigma$, are graded complex vector
spaces. The bundles $E^+$ and $E^-$ are identified with the $U(N^+)$ and
$U(N^-)$ Chan-Paton gauge bundles on the branes and antibranes, respectively.
When $N^+=N^-$, $E^+$ and $E^-$ are topologically the same, so that when such a
collection of branes annihilates to the vacuum state there is no overall
D-brane charge \cite{Sen1}. For $N^+\neq N^-$, after brane-antibrane
annihilation one is left with an excited state that has the Ramond-Ramond
charge of $N^+-N^-$ D-branes.

This means that the natural geometrical objects to consider on the
$p$-$\overline{p}$ worldvolume are not ordinary gauge connections, but rather
superconnections \cite{Quillen1}--\cite{BGV}. For this, we let
$\Omega(\Sigma)=\bigoplus_{k\geq0}\Omega^k(\Sigma)$ be the graded algebra of
smooth complex-valued differential forms over $\Sigma$ with $\zed$-grading
defined by the form degree $k$. The space $\Omega(\Sigma,E)$ of smooth
$E$-valued differential forms on $\Sigma$ then has a natural $\zed\times\zed_2$
grading, but we will be mainly concerned with its total $\zed_2$-grading
defined by $\Omega(\Sigma,E)=\Omega^+(\Sigma,E)\oplus\Omega^-(\Sigma,E)$, where
\beq
\Omega^\pm(\Sigma,E)=\bigoplus_{k\geq0}\left(\Omega^{2k}(\Sigma,E^\pm)
\oplus\Omega^{2k+1}(\Sigma,E^\mp)\right) \ .
\label{Omegagrad}\eeq
A superconnection is then any odd linear operator $\SA$ on the
$\Omega(\Sigma)$-module $\Omega(\Sigma,E)$, i.e.
$\SA:\Omega^\pm(\Sigma,E)\to\Omega^\mp(\Sigma,E)$, that satisfies the Leibnitz
rule
\beq
[\SA,\beta]^+=d\beta~~~~~~,~~~~~~\beta\in\Omega(\Sigma) \ ,
\label{Leibnitz}\eeq
where $[\cdot,\cdot]^+$ denotes the graded commutator. Note that a
superconnection does not necessarily send $k$-forms to $k+1$-forms, but rather
odd (resp. even) elements to even (resp. odd) elements. Nonetheless, because of
the Leibnitz property (\ref{Leibnitz}), the superconnections on a Chan-Paton
superbundle always form an affine space modelled on some set of local
operators.

If $\nabla$ is an arbitrary connection, then, by the Leibnitz rule,
$\SA-\nabla$ commutes with elements of $\Omega(\Sigma)$, and so it can be
represented by the exterior product with an odd matrix-valued form ${\cal A}$,
i.e. $(\SA-\nabla)\beta={\cal A}\wedge\beta$ for some ${\cal
A}\in\Omega^-(\Sigma,{\rm End}\,E)$. Thus any superconnection can be written in
terms of a fixed, fiducial connection $\nabla$ as
\beq
\SA=\nabla+{\cal A} \ .
\label{SAnablarel}\eeq
{}From the tensor product grading on the endomorphism algebra
$\Omega(\Sigma,{\rm End}\,E)$ we have
\bea
\Omega^\pm(\Sigma,{\rm End}\,E)&=&\left[\Omega^\mp(\Sigma)\otimes\left({\rm
Hom}(E^+,E^-)\oplus{\rm Hom}(E^-,E^+)\right)\right]\nn\\&
&\oplus\left[\Omega^\pm(\Sigma)\otimes\left({\rm End}\,E^+\oplus{\rm
End}\,E^-\right)\right] \ ,
\label{OmEndgrad}\eea
with the multiplication on $\Omega(\Sigma,{\rm End}\,E)=\Omega^+(\Sigma,{\rm
End}\,E)\oplus\Omega^-(\Sigma,{\rm End}\,E)$ given by
\beq
(\alpha\otimes a)\cdot(\beta\otimes
b)=(-1)^{|a||\beta|}\,(\alpha\wedge\beta)\otimes(a\circ b)
\label{OmEndmult}\eeq
for $\alpha,\beta\in\Omega(\Sigma)$ and $a,b\in{\rm End}\,E$, where $|a|$
denotes the total degree of $a$. With respect to this $\zed_2$-grading, we may
decompose the linear operator $\cal A$ as
\beq
{\cal A}=\pmatrix{A^+&T^-\cr T^+&A^-\cr}
\label{calAmatrixgen}\eeq
where
\bea
A^\pm&=&\sum_{k\geq0}A^\pm_{(2k+1)}\in\Omega^-(\Sigma)\otimes{\rm End}\,E^\pm
 \ , \nn\\T^\pm&=&\sum_{k\geq0}T_{(2k)}^\pm\in\Omega^+(\Sigma)\otimes
{\rm Hom}(E^\pm,E^\mp) \ .
\label{ATexpansion}\eea
The one-form components of $A^\pm$ define ordinary gauge connections
$A_{(1)}^\pm$ on the bundles $E^\pm$. The zero-form components of $T^\pm$ are
odd matrix-valued bundle maps $T^\pm_{(0)}: C^\infty(\Sigma,E^\pm)\to
C^\infty(\Sigma,E^\mp)$ with $(T^\pm_{(0)})^\dagger=T^\mp_{(0)}$, i.e. smooth
sections of the product bundles $E^\mp\otimes(E^\pm)^*$, where $(E^\pm)^*$ is
the dual of $E^\pm$. They define the complex scalar tachyon fields
$T(x)=T^+_{(0)}(x)$ of the brane-antibrane system. At a point $x\in\Sigma$, the
tachyon field is a linear fiber map $T(x):E_x^+\to E_x^-$ and its adjoint
$T^\dagger(x):E_x^-\to E_x^+$.

If $\theta$ is any matrix-valued form, then the Leibnitz rule and the Jacobi
identity for the graded commutator imply that
\beq
\left[[\SA,\theta]^+\,,\,\beta\right]^+=\left[\SA\,,\,[\theta,\beta]^+\right]^+
+(-1)^{|\theta||\beta|}\,\Bigl[d\beta\,,\,\theta\Bigr]^+~~~~~~
\forall\beta\in\Omega(\Sigma) \ .
\label{JacLeib}\eeq
Eq.~(\ref{JacLeib}) defines the action of the covariant derivative
$\SA\,\theta$ in $\Omega(\Sigma,{\rm End}\,E)$, i.e. $\SA\,\theta$ is
identified as multiplication by the operator $[\SA,\theta]^+$. Since $\SA$ is
odd, we have $[\SA,\SA]^+=2\SA^2$, and so from the Leibnitz rule and
(\ref{JacLeib}) we find
\beq
\Bigl[\SA^2\,,\,\beta\Bigr]^+=\left[\SA\,,\,[\SA,\beta]^+\right]^+=d(d\beta)=0
{}~~~~~~\forall\beta\in\Omega(\Sigma) \ .
\label{SA2comm}\eeq
It follows that $\SA^2$ lives in the endomorphism bundle $\Omega^+(\Sigma,{\rm
End}\,E)$. The operator $F_\sSA=\SA^2$ is the curvature of the superconnection
$\SA$ and it satisfies the Bianchi identity
\beq
\SA F_\sSA=\Bigl[\SA\,,\,F_\sSA\Bigr]^+=\left[\SA\,,\,\SA^2\right]^+=0 \ .
\label{Bianchi}\eeq
The curvature is in general a sum
\beq
F_\sSA=\sum_{k\geq0}{\cal F}_{(k)}~~~~~~,~~~~~~{\cal
F}_{(k)}\in\Omega^k(\Sigma,{\rm End}\,E) \ .
\label{FAcalFk}\eeq
By using the grading (\ref{OmEndgrad}) the first few components are found to be
\bea
{\cal F}_{(0)}&=&\pmatrix{T^\dagger T&0\cr0&TT^\dagger\cr} \ ,
\label{TTdagger}\\{\cal F}_{(1)}&=&\pmatrix{0&DT^\dagger\cr DT&0\cr} \ ,
\label{covderivT}\\{\cal F}_{(2)}&=&\pmatrix{R_\nabla+F^+&0\cr
0&R_\nabla+F^-\cr} \ , \label{ordcurvs}\\&\vdots&\nn
\eea
where $R_\nabla=\nabla^2$ is the curvature of the fiducial derivation $\nabla$,
\beq
F^\pm=F_{A^\pm}=\nabla A^\pm_{(1)}+A_{(1)}^\pm\wedge A_{(1)}^\pm
\label{Fpm}\eeq
are the curvatures of the gauge connections $A_{(1)}^\pm$ on $E^\pm$, and
\bea
DT&=&\nabla T+A_{(1)}^-\,T+TA_{(1)}^+ \ , \nn\\
DT^\dagger&=&\nabla T^\dagger+A_{(1)}^+\,T^\dagger+T^\dagger A_{(1)}^- \ .
\label{DTdef}\eea
The curvature component (\ref{TTdagger}) defines a term bilinear in the tachyon
field, the component (\ref{covderivT}) defines a kinetic term in terms of
covariant derivatives of the tachyon field, and (\ref{ordcurvs}) yields the
usual Yang-Mills field strengths on the branes and antibranes in the absence of
the tachyon field (and all other higher rank fields).

Everything we have said thus far has been rather general, and we now need to
input some more physical requirements. We will focus on the low-lying
excitations of the brane-antibrane system, which are described by the $p$-$p$,
$p$-$\overline{p}$ and $\overline{p}$-$\overline{p}$ open string states. The
$p$-$p$ (resp. $\overline{p}$-$\overline{p}$) open string spectrum consists of
a massless $U(N^+)$ (resp. $U(N^-)$) supersymmetric Yang-Mills multiplet, along
with massive excitations. In these bosonic components the NS sector tachyon
state is removed by the standard GSO projection. The open string wavefunctions
are the products $\psi=\psi_{\rm osc}\otimes\psi^{~}_{\rm CP}$ of the usual
mode decompositions and the Chan-Paton factors $\psi^{~}_{\rm CP}\in
U(N^++N^-)$. The GSO projection operator is
\beq
P_{\rm GSO}=\frac12\,\Bigl(\id+(-1)^F\Bigr)
\label{PGSO}\eeq
where $F$ is the worldsheet fermion number operator. Its action on the
Chan-Paton factors may be represented as
\beq
(-1)^F\,:\,\psi^{~}_{\rm CP}~\longmapsto~\varepsilon\,\psi^{~}_{\rm
CP}\,\varepsilon
\label{gradingCP}\eeq
in terms of the usual grading automorphism $\varepsilon$ for the superbundle
$E$,
\beq
\varepsilon=\pmatrix{\id_{N^+}&0\cr0&-\id_{N^-}\cr} \ ,
\label{gradingop}\eeq
such that $E=E^+\oplus E^-$ decomposes into the $\pm1$ eigenspaces of
$\varepsilon$. With respect to this decomposition, the $p$-$p$ and
$\overline{p}$-$\overline{p}$ open strings have diagonal Chan-Paton
wavefunctions $\psi^{~}_{\rm CP}$ which are even under the action
(\ref{gradingCP}) of $(-1)^F$, leading to the usual GSO projection on the
oscillator modes $\psi_{\rm osc}$. On the other hand, the Chan-Paton
wavefunctions for the $p$-$\overline{p}$ and $\overline{p}$-$p$ open string
states are off-diagonal and odd under $(-1)^F$, leading to a reversed GSO
projection on the corresponding oscillators. This means that in these sectors
the massless $U(N^+)$ and $U(N^-)$ vector supermultiplets are projected out,
and the tachyon survives \cite{BanksSuss,GG1}. These features of the low-energy
open string theory are already encoded in the first two components of the
superconnection constructed above. But it also implies that, as far as the
low-energy effective field theory on the brane-antibrane worldvolume is
concerned, all higher form degree components of $\cal A$ in
(\ref{calAmatrixgen}) are absent, because the GSO projection (\ref{PGSO})
eliminates the off-diagonal fermionic gauge fields. The superconnections
relevant to the low-energy physics of brane-antibrane systems are therefore
precisely of the type considered originally in \cite{Quillen1}, and henceforth
we shall thereby deal only with the superconnection defined by taking
$A^\pm=A_{(1)}^\pm$ and $T^\pm=T^\pm_{(0)}$ in (\ref{calAmatrixgen}). Then, the
superconnection field strength (\ref{FAcalFk}) reduces to
\beq
F_\sSA=\pmatrix{R_\nabla+F^++T^\dagger T&DT^\dagger\cr
DT&R_\nabla+F^-+TT^\dagger\cr} \ .
\label{FAred}\eeq

The stated properties of the GSO projection also imply that the gauge symmetry
group $G$ of the superbundle $E=E^+\oplus E^-$, which is generically the
unitary Lie supergroup $U(N^+|N^-)$, is instead that which is lifted from the
structure groups of the Chan-Paton bundles $E^\pm$ over the branes and
antibranes, i.e. $G=U(N^+)\times U(N^-)$.\footnote{\baselineskip=12pt An
alternative interpretation of the $U(N^+|N^-)$ brane-antibrane supergroup
symmetry in the context of topological string theory is given in~\cite{Vafa}.}
With $g=g^+\oplus g^-$ an automorphism of the associated principal bundle, the
gauge transformation law of the superconnection is given as
\bea
{\cal A}&\longmapsto&g\,{\cal A}\,g^{-1}+g\,\nabla g^{-1} \ , \nn\\
F_\sSA&\longmapsto&g\,F_\sSA\,g^{-1} \ ,
\label{supergaugetransf}\eea
which is equivalent to the component transformation rules
\bea
A_{(1)}^\pm&\longmapsto&g^\pm\,A_{(1)}^\pm\,(g^\pm)^{-1}+g^\pm\,
\nabla(g^\pm)^{-1} \ , \nn\\
F^\pm&\longmapsto&g^\pm\,F^\pm\,(g^\pm)^{-1} \ , \nn\\
T&\longmapsto&g^-\,T\,(g^+)^{-1} \ , \nn\\
T^\dagger&\longmapsto&g^+\,T^\dagger\,(g^-)^{-1} \ .
\label{compgaugetransf}\eea
In other words, the component gauge fields transform in the usual way under the
adjoint actions of $U(N^+)$ and $U(N^-)$, while the tachyon field transforms in
a bi-fundamental unitary group representation $\overline{{\bf N}^-}\otimes{\bf
N}^+$, or equivalently it carries charges $(\overline{{\bf N}^-},{\bf N}^+)$
with respect to the $U(N^-)\times U(N^+)$ brane-antibrane gauge fields
$(A_{(1)}^-,A_{(1)}^+)$. But within the more general superconnection formalism
presented above, we see that the tachyon field may actually be regarded as a
sort of generalized gauge field on the discrete space $\zed_2$, i.e. we may
regard the brane-antibrane worldvolume as the two-sheeted manifold
$\Sigma\times\zed_2$, whose two sheets are connected together by $T$.

\subsection{Spin Geometry}

The properties of the GSO projection on the low-lying excitations of the
brane-antibrane system have in addition some important consequences for the
spin geometry of the worldvolume manifold $\Sigma$. We work in Type IIB
superstring theory, so that $\dim\Sigma=p+1$ is even, and assume that $\Sigma$
is oriented with a given Riemannian structure. The corresponding result for
Type IIA D-branes will follow by T-duality. Let $\cliff(\Sigma)$ be the complex
Clifford bundle over $\Sigma$ whose fiber over a point $x\in\Sigma$ is the
complexified Clifford algebra $\cliff(T_x^*\Sigma)$ with respect to a Hermitian
structure $\langle\cdot,\cdot\rangle$ on the fibers of the cotangent bundle
$T^*\Sigma$. The smooth sections of the Clifford bundle form the algebra ${\cal
C}= C^\infty(\Sigma,\cliff\,\Sigma)$. Let $S\to\Sigma$ be a spinor bundle of
rank $2^{(p+1)/2}$, and let $c:T^*\Sigma\to{\rm End}\,S$ be a
spin$^c$-structure on the worldvolume $\Sigma$. This requires, in addition to
the orientability of $\Sigma$, that its second Stiefel-Whitney class
$w_2(\Sigma)$ be the mod 2 reduction of an integral cohomology class. The
linear bundle map $c$ satisfies
\beq
c(v)^2+\langle v,v\rangle=0~~~~~~,~~~~~~v\in T^*\Sigma \ ,
\label{spinceq}\eeq
and it can be universally extended to an irreducible Clifford action on $S$,
i.e. $c$ extends uniquely to an algebra isomorphism
$c:\cliff(\Sigma)\stackrel{\approx}{\longrightarrow}{\rm End}\,S$ which is
compatible with the property (\ref{spinceq}). In particular, the algebra $\cal
C$ acts irreducibly on the space $ C^\infty(\Sigma,S)$ of smooth sections of
the spinor bundle, so that ${\cal C}\cong{\rm End}\,S$.

The Clifford bundle is a superbundle
$\cliff(\Sigma)=\cliff^+(\Sigma)\oplus\cliff^-(\Sigma)$ with grading
automorphism $v\mapsto-v$. The spinor bundle is also a superbundle $S=S^+\oplus
S^-$ with $\zed_2$-grading defined as follows. For each $x\in\Sigma$, let
$\theta^a$ be an oriented orthonormal frame in $T_x^*\Sigma$, and set
\beq
\gamma^a=c(\theta^a) \ .
\label{gammaa}\eeq
The $\gamma$'s generate locally the Euclidean Dirac algebra
\beq
\gamma^a\gamma^b+\gamma^b\gamma^a=-2\delta^{ab}~~~~~~a,b=1,\dots,p+1 \ .
\label{Diracalg}\eeq
The Hermitian chirality operator $\gamma_{\rm
c}=i^{(p+1)/2}\,\gamma^1\gamma^2\cdots\gamma^{p+1}$ then satisfies
\bea
(\gamma_{\rm c})^2&=&\id_{2^{(p+1)/2}} \ , \nn\\
\gamma_{\rm c}\,\gamma^a&=&-\gamma^a\,\gamma_{\rm c} \ .
\label{gammacprops}\eea
The sub-bundles $S^\pm$ are taken to be the $\pm1$ eigenspaces of the chirality
operator $\gamma_{\rm c}$. They have the same rank $2^{(p-1)/2}$, and because
of (\ref{gammacprops}), the (left) action of $\cliff(\Sigma)$ on $S$, defined
by $a\cdot s=c(a)s$, preserves this $\zed_2$-grading, i.e.
$\cliff^+(\Sigma)\cdot S^\pm\subset S^\pm$ and $\cliff^-(\Sigma)\cdot
S^\pm\subset S^\mp$. Thus the superbundle $S=S^+\oplus S^-$ is a graded (left)
Clifford module.

We will now incorporate the graded Chan-Paton vector bundle $E=E^+\oplus E^-$
of the brane-antibrane system above. For this, we introduce the twisted spinor
bundle
\beq
S_E=S\otimes E=S_E^+\oplus S_E^-
\label{twistspin}\eeq
with Clifford action $c\otimes\id$, which we also denote by $c$. From the
properties of the GSO projection described in the previous subsection, it
follows that the only chiral spinors that appear in the spectrum of the
brane-antibrane system are those which are lifted over the bundles $E^\pm$. The
GSO projection removes the massless fermionic modes in the open string
$p$-$\overline{p}$ and $\overline{p}$-$p$ sectors. This implies that the
$\zed_2$-grading of (\ref{twistspin}) is just that which arises from the
induced tensor product grading of the two superbundles $S$ and $E$,
\beq
S_E^\pm=\left(S^\pm\otimes E^+\right)\oplus\left(S^\mp\otimes E^-\right) \ .
\label{twistspingrad}\eeq
This makes $S_E$ a Clifford module. It also implies that the appropriate
superconnection on (\ref{twistspin}) is a Clifford superconnection $\SS$, i.e.
$\SS$ respects the $\cal C$-module structure by satisfying a second Leibnitz
rule (in addition to (\ref{Leibnitz})) involving the Clifford action,
\beq
\Bigl[\SS\,,\,c(\beta)\Bigr]^+=c(\nabla\beta)~~~~~~\forall\beta\in\Omega(\Sigma) \ .
\label{Leibnitz2}\eeq
Here $\nabla: C^\infty(\Sigma,T\Sigma)\to C^\infty(\Sigma,T^*\Sigma\otimes
T\Sigma)$ is the Levi-Civita connection on the tangent bundle $T\Sigma$ which
can be written locally in terms of Christoffel symbols
$\Gamma\in\Omega^1(\Sigma,so(T^*\Sigma))$ as $\nabla=d+\Gamma$.

The spin connection $\nabla^{\rm s}$ on the spinor bundle has the property
(\ref{Leibnitz2}). Because of the canonical bundle isomorphism
$\Omega(\Sigma)\cong\cliff(\Sigma)$, it is a map $\nabla^{\rm s}:
C^\infty(\Sigma,S)\to C^\infty(\Sigma,T^*\Sigma\otimes S)$ which can be written
locally as
\beq
\nabla^{\rm s}=d+\omega(\Gamma) \ ,
\label{spinconnloc}\eeq
where $\omega:so(T^*\Sigma)\to\cliff(\Sigma)$ is the spinor representation of
the Lie algebra of the orthogonal group with
\beq
\omega(\Gamma_i)=-\frac14\,\Gamma_{i\,a}^b\,\gamma^a\gamma^b \ .
\label{spinrepinf}\eeq
The corresponding curvature is
\beq
(\nabla^{\rm
s})^2=\omega(R_\nabla)=\frac14\,R_{baij}\,\gamma^a\gamma^b\,dx^i\wedge dx^j
\label{spincurv}\eeq
where $R_\nabla=d\Gamma+\Gamma\wedge\Gamma\in\Omega^2(\Sigma,so(T^*\Sigma))$ is
the Riemann curvature tensor. It follows that if $\SS$ is any Clifford
superconnection on $S_E$, then $\SS-\nabla^{\rm s}\otimes\id$ commutes with the
Clifford action $c$, and therefore the most general Clifford superconnection is
of the form
\beq
\SS=\nabla^{\rm s}\otimes\id+\id\otimes\SA
\label{cliffsuperconn}\eeq
where $\SA$ is any superconnection on the twisting bundle $E$. Since
$[\nabla^{\rm s}\otimes\id,\id\otimes\SA]=0$, the curvature of
(\ref{cliffsuperconn}) splits as
\beq
\SS^2=\omega(R_\nabla)\otimes\id+\id\otimes F_\sSA
\label{superconncurv}\eeq
into the sum of a purely gravitational piece and an internal gauge piece. Thus
the Clifford superconnections $\SS$ will naturally incorporate both gauge and
gravitational effects into the ensuing computations. We note that
(\ref{superconncurv}) is compatible with the identification
$F_\sSA\in\Omega^+(\Sigma,{\rm End}\,E)$, because the Leibnitz rule
(\ref{Leibnitz2}) implies
\beq
\Bigl[\SS^2\,,\,c(\beta)\Bigr]^+=\left[\SS\,,\,\Bigl[\SS\,,\,c(\beta)\Bigr]^+
\right]^+=c\left(\nabla^2\beta\right)=c\left(R_\nabla\wedge\beta\right)
\label{SS2curv}\eeq
and also $[\omega(R_\nabla),c(\beta)]^+=c(R_\nabla\wedge\beta)$. This shows
that $F_\sSA=\SS^2-\omega(R_\nabla)$ commutes with all $c(\beta)$, as expected.

Thus, given any superconnection $\SA$ on a Chan-Paton superbundle $E=E^+\oplus
E^-$, there is a natural extension to a Clifford superconnection
(\ref{cliffsuperconn}) on the twisted spinor bundle (\ref{twistspin}). In some
applications, such as the bound state constructions of lower dimensional
D-branes from the brane-antibrane system, we will require that the Chan-Paton
bundle $E$ itself define a Clifford module. Then Schur's lemma implies that any
map $ C^\infty(\Sigma,S)\to C^\infty(\Sigma,E)$ which commutes with the
Clifford action $c$ is of the form $\psi\mapsto\psi\otimes w$, where $w\in
C^\infty(\Sigma,W)$ with $W={\rm Hom}_{\cliff\,\Sigma}(S,E)$ the bundle of
intertwining maps \cite{BGV}. Moreover, any endomorphism of $
C^\infty(\Sigma,E)$ that commutes with the Clifford action is of the form
$\psi\otimes w\mapsto\psi\otimes Lw$ for some bundle map $L:W\to W$, so that
${\rm End}_{\cliff\,\Sigma}\,E\cong\id\otimes{\rm End}\,W$. The entire matrix
bundle ${\rm End}\,E$ is then generated by the sub-bundle
$\cliff(\Sigma)\cong{\rm End}\,S\otimes\id$ acting by the spinor
representation, and by its commutant $\id\otimes{\rm End}\,W$, so that
\beq
{\rm End}\,E\cong\cliff(\Sigma)\otimes{\rm End}\,W \ .
\label{EndEcliff}\eeq
This means that any Clifford module $ C^\infty(\Sigma,E)$ on a
spin$^c$-manifold $\Sigma$ comes from a twisted spinor bundle $E\cong
S_W=S\otimes W$ with the canonical grading defined by (\ref{twistspingrad}). A
generic Clifford superconnection $\SA$ on $E$ may then be written as
$\SA=\nabla^{\rm s}\otimes\id+\id\otimes\SW$, where $\SW$ is any
superconnection on the intertwining bundle $W$.

\subsection{Generalized Dirac Operators}

We will now describe some general properties of Dirac operators associated with
superconnections that are compatible with the Clifford action, with respect to
a chosen spin$^c$-structure on $\Sigma$. These operators will play a
fundamental role in all considerations of this paper. Let $\SS$ be such a
Clifford superconnection, as in (\ref{cliffsuperconn}). We identify the algebra
${\cal C}= C^\infty(\Sigma,\cliff\,\Sigma)$ with the algebra of differential
forms $\Omega(\Sigma)$ by the symbol map isomorphism
$\sigma(\beta)=c(\beta)\,\id_S$. The inverse ``quantization'' map
$Q:\Omega(\Sigma)\stackrel{\approx}{\longrightarrow}
C^\infty(\Sigma,\cliff\,\Sigma)$ then allows one to represent exterior products
of forms by Clifford multiplication \cite{BGV}. We define the associated
twisted Dirac operator by the following composition of maps,
\beq
\Sdirac\,:~ C^\infty(\Sigma,S_E)~\stackrel{\sSS}{\longrightarrow}~
C^\infty(\Sigma,
T^*\Sigma\otimes S_E)~\stackrel{c}{\longrightarrow}~ C^\infty(\Sigma,S_E) \ .
\label{Diracmaps}\eeq
In local coordinates where $\SS=\theta^a\otimes\SS_a$ we have
$\Sdirac=\gamma^a\,\SS_a$, and with respect to the grading (\ref{OmEndgrad}) we
may write
\beq
\Sdirac=\pmatrix{\dirac_A^+&{\cal G}_{\rm s}\otimes T^\dagger\cr
{\cal G}_{\rm s}\otimes T&\dirac_A^-\cr}
\label{Diracmatrix}\eeq
where
\beq
\dirac_A^\pm=Q\left(\nabla^{\rm s}\otimes\id+\id\otimes
A_{(1)}^\pm\right)=\gamma^a\,\nabla^{\rm s}_a\otimes\id+\gamma^a\otimes
A_{(1)\,a}^\pm \ .
\label{DiracQ}\eeq
Here ${\cal G}_{\rm s}={\cal G}_{\rm s}^\dagger$ is a constant operator on $
C^\infty(\Sigma,S)$ which we have combined with the tachyon field in the
zero-form component of the superconnection $\SA$.

The Dirac operator $\Sdirac$ on $ C^\infty(\Sigma,S_E)$ satisfies three
fundamental conditions, namely
\begin{itemize}
\item{It is an odd operator, i.e. $\Sdirac: C^\infty(\Sigma,S_E^\pm)\to
C^\infty(\Sigma,S_E^\mp)$.}
\item{It respects the $\cal C$-module structure of $ C^\infty(\Sigma,S_E)$,
i.e. $\Sdirac$ satisfies (\ref{Leibnitz2}).}
\item{It transforms homogeneously under local gauge transformations
(\ref{supergaugetransf},\ref{compgaugetransf}), i.e. $\Sdirac\mapsto
g\,\Sdirac\,g^{-1}$.}
\end{itemize}
In particular, there is a one-to-one correspondence between Dirac operators
which are compatible with a given Clifford action and Clifford superconnections
\cite{BGV}, so that $\Sdirac$ and $\SS$ carry the same information. In this
sense, superconnections may be thought of as quantizations of ordinary
connections \cite{Quillen1}.

\subsection{Transgression and Descent Equations}

Most of our subsequent analysis will rely on the computation of gauge anomalies
associated with the brane-antibrane system. In this subsection we will show how
the standard formalism for Yang-Mills anomalies carries over to the case of
superconnection gauge fields. For this, we consider an arbitrary function
$I(\alg,F_\sSA)$ of a superconnection $\SA$ and its curvature $F_\sSA$. We
introduce the integral operator
\beq
\SK\,I(\alg,F_\sSA)=\int\limits_0^1dt~\SK_{(t)}\,
I\Bigl(t\alg\,,\,F_\sSA(t)\Bigr) \ ,
\label{SKdef}\eeq
where $F_\sSA(t)=t\,d\alg+t^2\alg\wedge\alg$ is the curvature associated with
$t\alg$ which is a path in field space connecting the superconnection gauge
fields 0 and $\alg$, and the anti-differential operator $\SK_{(t)}$ is defined
by $\SK_{(t)}(t\alg)=0$, $\SK_{(t)}\,F_\sSA(t)=t\alg$. From this definition we
can infer the generalization of the Cartan homotopy formula
\beq
I(\alg,F_\sSA)=\left(\SK\,d+d\,\SK\right)I(\alg,F_\sSA) \ .
\label{Cartanformula}\eeq

We will now specialize to the case of the invariant polynomial
$I(\alg,F_\sSA)=\Tr^+(F_\sSA)^n$, where $n\geq1$ and
$\Tr^+(\cdot)=\Tr(\varepsilon\,\cdot)$ is the supertrace in the fundamental
representation of the brane-antibrane gauge group $U(N^+)\times U(N^-)$. This
polynomial is a closed form, because the Bianchi identity (\ref{Bianchi})
implies
\beq
d\,\Tr^+(F_\sSA)^n=\Tr^+\left(d(F_\sSA)^n+\Bigl[\alg\,
\stackrel{\wedge}{,}\,(F_\sSA)^n\Bigr]^+\right)=0 \ .
\label{closedpoly}\eeq
The important property of this closed differential form is that its cohomology
class is independent of the choice of superconnection $\SA$ \cite{Quillen1}. To
see this, let $\SA_s$, $s\in\real$, be a one-parameter family of
superconnections with curvatures $F_s=(\SA_s)^2$. Then
\beq
\frac{\partial F_s}{\partial s}=\left[\SA_s\,,\,\frac{\partial\SA_s}
{\partial s}\right]^+ \ ,
\label{dFsds}\eeq
and applying this result to the invariant polynomial $\Tr^+(F_s)^n$ gives
\bea
\frac{\partial\,\Tr^+(F_s)^n}{\partial s}&=&n\,\Tr^+\,\frac{\partial F_s}
{\partial s}\,\wedge(F_s)^{n-1}\nn\\
&=&n\,\Tr^+\left[\SA_s\,,\,\frac{\partial \SA_s}{\partial s}
\,(F_s)^{n-1}\right]^+\nn\\
&=&d\,\Tr^+\,\frac{\partial\SA_s}{\partial s}\,(F_s)^{n-1} \ ,
\label{superconndeform}\eea
showing that any continuous deformation of the form $\Tr^+(F_\sSA)^n$ changes
it by an exact form. In particular, the cohomology class determined by it is
independent of the choice of profile for the tachyon field $T$, a fact which
will be exploited throughout this paper.

{}From (\ref{closedpoly}) we arrive at the generalized transgression formula
\cite{CS}
\beq
\Tr^+(F_\sSA)^n=d\xi_{2n-1}^{(0)}(\alg,F_\sSA) \ ,
\label{transgression}\eeq
where $\xi_{2n-1}^{(0)}$ is a generalized Chern-Simons form which using the
homotopy formula (\ref{Cartanformula}) can be written as
\beq
\xi_{2n-1}^{(0)}(\alg,F_\sSA)=\SK\,\Tr^+(F_\sSA)^n=n\int\limits_0^1dt~
\Tr^+\,\alg\wedge F_\sSA(t)^{n-1} \ .
\label{CS0}\eeq
The form (\ref{CS0}) is the first member of the BRST complex constructed from
the Lie algebra cohomology of $U(N^+)\times U(N^-)$. Let $\delta_{\rm BRST}$ be
the corresponding coboundary operator which is associated with the
infinitesimal gauge transformations (\ref{supergaugetransf}). The BRST ghost
field $\Lambda=\Lambda^+\oplus\Lambda^-$ is the Cartan-Maurer one-form on the
brane-antibrane gauge group, i.e. the tautological one-form
\beq
\Lambda=g^{-1}\,\delta_{\rm BRST}\,g
\label{Cartanform}\eeq
at the identity element which sends a Lie algebra element onto itself. It
satisfies the Cartan-Maurer equation
\beq
\delta_{\rm BRST}\,\Lambda=-\Lambda\wedge\Lambda \ .
\label{CMeqn}\eeq
The other terms in the BRST complex are then obtained by replacing the
superconnection via $\alg\mapsto\alg+\Lambda$ and with the curvature associated
with the operator $d+\delta_{\rm BRST}$. The appropriate generalizations of
(\ref{CS0}) then come from the paths of superconnections in field space which
connect $\Lambda$ with $\alg+\Lambda$, and also 0 with $\Lambda$. They
respectively read
\beq
\xi_{2n-1-k}^{(k)}=\left\{\new{\begin{array}{l}
n\int\limits_0^1dt~\Tr^+\left[\alg\wedge
\Bigl(F_\sSA(t)+(1-t)\,d\Lambda\Bigr)^{n-1}
\right]^{(k)}~~~~~~{\rm for}~~k=0,1,\dots,n-1 \ , \\
n\int\limits_0^1dt~\Tr^+\left[\Lambda\wedge\Bigl(t\,d\Lambda+(t^2-t)\,
\Lambda\wedge\Lambda\Bigr)^{n-1}
\right]^{(k)}~~~~~~{\rm for}~~k=n,n+1,\dots,2n \ , \end{array}}\right.
\label{CSk}\eeq
where the bracket $[\,\cdot\,]^{(k)}$ indicates to extract the terms of degree
$k$ in the BRST ghost field~$\Lambda$.

The forms (\ref{CSk}) are related to one another through the descent equations
\cite{WZ}
\beq
\delta_{\rm BRST}\,\xi_{2n-1-k}^{(k)}=-d\xi_{2n-k-2}^{(k+1)}~~~~~~,~~~~~~
k=0,1,\dots,2n-1 \ .
\label{descenteqs}\eeq
The cocycles $\xi_{2n-1-k}^{(k)}$ may also be obtained in a somewhat more
geometric way via the Cheeger-Simons construction~\cite{CheegerSimons}. For
this, we fix a superconnection on the graded universal bundle $EG\to BG$, where
$BG$ is a smooth classifying space of the brane-antibrane gauge group
$G=U(N^+)\times U(N^-)$. Let $\widehat{\xi}_{2n-1-k}^{\,(k)}$ be a cocycle
representative of the Chern-Weil form associated to a Chern-Simons
characteristic class in $H^{2n-1-k}(BG,\real)$. Given a Chan-Paton superbundle
$E\to\Sigma$ with superconnection $\SA$, let $f:\Sigma\to BG$ be the map
induced from a homotopy class of $G$-equivariant classifying maps on $E\to EG$.
Then the Chern-Simons cocycle of $\SA$ is given by the pullback
\beq
\xi_{2n-1-k}^{(k)}=f^*\widehat{\xi}^{\,(k)}_{2n-1-k}\in
H^{2n-1-k}(\Sigma,\real) \ .
\label{CScocycles}\eeq
By suitably integrating the forms (\ref{CScocycles}) over cycles of the
worldvolume $\Sigma$, one obtains a family of cocycles ${\rm
CS}_{2n-1-k}^{(k)}$, $\delta_{\rm BRST}\,{\rm CS}_{2n-1-k}^{(k)}=0$, in the
BRST cohomology $H^k(U(N^+)\times U(N^-),\real)$ of the brane-antibrane gauge
group. In what follows we shall deal mostly with the topological anomaly ${\rm
CS}_{2n-2}^{(1)}\in H^1(U(N^+)\times U(N^-),\real)$ which corresponds to the
$k=1$ term in (\ref{descenteqs}). The form
\beq
\xi_{2n-2}^{(1)}=n(n-1)\int\limits_0^1dt~(1-t)\,{\rm STr}^+\,\Lambda\wedge
d\Bigl(\alg\wedge F_\sSA(t)^{n-2}\Bigr)
\label{chiralform}\eeq
is then the solution to the Wess-Zumino consistency condition \cite{WZ} for the
chiral gauge anomaly in the brane-antibrane worldvolume field theory, where
\beq
{\rm STr}(M_1,\dots,M_n)=\frac1{n!}\,\sum_{\pi\in S_n}
\Tr\left(M_{\pi_1}\cdots M_{\pi_n}\right)
\label{symtrace}\eeq
is the symmetrized trace. For example, setting $n=2$ in the above formulas
yields the generalized Chern-Simons form
\bea
\xi_3^{(0)}(\alg,F_{\sSA})&=&\Tr\left(A_{(1)}^+\wedge dA_{(1)}^++
\frac23\,A_{(1)}^+\wedge A_{(1)}^+\wedge A_{(1)}^+\right)\nn\\&&
-\,\Tr\left(A_{(1)}^-\wedge dA_{(1)}^-+
\frac23\,A_{(1)}^-\wedge A_{(1)}^-\wedge A_{(1)}^-\right)+\Tr\left(
T\,DT^\dagger+T^\dagger\,DT\right) \ . \nn\\&&
\label{CS03}\eea
The first two terms in (\ref{CS03}) are the standard three-dimensional
Chern-Simons terms for the gauge fields on the branes and antibranes,
respectively. The last term represents the modification of the ordinary anomaly
formula due to the tachyon field.

The analysis of this subsection shows that the standard techniques for dealing
with gauge anomalies carry through to the case of superconnections, with the
appropriate modifications. It is also possible to incorporate modifications due
to the gravitational terms on the brane-antibrane system by constructing
superconnections from the generalized Dirac operators of the previous
subsection. We shall return to this point in section~6. Brane actions based
on transgression forms have been constructed from a different point of view
in~\cite{Mora}.

\newsection{Ramond-Ramond Couplings on Brane-Antibrane Systems}

In this section we will apply the previous considerations to a systematic
derivation of the Ramond-Ramond couplings of brane-antibrane pairs. Amongst
other objects, the brane-antibrane worldvolume $\Sigma$ has spinor fields
$\Psi$ defined on it with kinetic term of the form
$\overline{\Psi}\,i\Dirac\,\Psi$ in the total Lagrangian of the worldvolume
field theory. If the fermion fields are chiral, then the functional integral
over them yields the regularized determinant of the (generalized) Dirac
operator $\Dirac$. By incorporating supergravity couplings to the D-branes, the
effective path integral measure for the brane-antibrane system will therefore
contain the factor
\beq
\Theta=\det(i\Dirac)~\e^{iZ} \ .
\label{Thetaanomdef}\eeq
Here
\beq
Z=-\frac{\mu}2\,\int\limits_{\Sigma}{\cal C}\wedge{\cal Y}
\label{WZansatz}\eeq
where $\mu$ is a charge associated with the branes, ${\cal
C}=\phi^*C=\sum_p\phi^*C_{(p)}$ is the pullback to
$\Sigma\stackrel{\phi}{\hookrightarrow}X$ of the total RR
potential,\footnote{\baselineskip=12pt We will assume throughout that the
worldvolume $\Sigma$ is embedded in the spacetime $X$. While this requirement
is not completely necessary, it will simplify some of the calculations in what
follows.} and the coupling ${\cal Y}$ will be determined in what follows as an
invariant polynomial of the gravitational and gauge curvatures on $\Sigma$. The
determinant in (\ref{Thetaanomdef}) depends on the various bosonic fields which
couple to $\Psi$ and which live in a parameter space $\cal M$. In many
instances it is not a function on $\cal M$ but rather a section of a complex
line bundle ${\cal P}(\Dirac)\to{\cal M}$ \cite{Freed1}. For chiral fermion
fields there is an obstruction to constructing a gauge invariant regularized
determinant of the Dirac operator $\Dirac$. It is the topological chiral
anomaly which is measured by the first Chern class of the determinant line
bundle over $\cal M$. A two-form representative for this Chern class can be
constructed as the transgression of the one-form ${\rm CS}_{2n-2}^{(1)}$. If
$\hil^+$ denotes the positive energy spectral subspace of $i\Dirac$, then the
cohomology class $[\hil^+]\in H^{2n-1}({\cal M},\zed)$ measures the
corresponding obstruction. The chiral gauge anomaly is the obstruction to
constructing a corresponding global, non-zero trivializing section on $\cal
M\to{\cal P}(\Dirac)$ which would allow the remaining functional integration
over $\cal M$ to be carried out in the brane-antibrane quantum field theory.
The topology of the determinant line bundle ${\cal P}(\Dirac)$ is given by the
Atiyah-Singer index theorem \cite{AS2}, which when expressed in differential
geometric terms gives explicit forms for the chiral anomaly \cite{AGDW}.

On the other hand, the action (\ref{WZansatz}) is not in general invariant
under local gauge transformations of the Ramond-Ramond tensor potentials. Its
exponential in (\ref{Thetaanomdef}) is therefore not a function on ${\cal
M}\to{\bf S}^1$, but rather a unit norm section of a complex line bundle over
the parameter space $\cal M$ which is not necessarily covariantly constant. To
make the brane-antibrane worldvolume quantum field theory well-defined, we will
demand that the product of sections in (\ref{Thetaanomdef}) be a globally
well-defined function on $\cal M$, i.e. that $\e^{-iZ}$ be a trivializing
section for $\det i\Dirac$. This will uniquely fix the form of the
Ramond-Ramond coupling (\ref{WZansatz}) to the brane-antibrane system. As in
the previous section, we shall throughout assume that the worldvolume manifold
$\Sigma$ admits a spin$^c$-structure, which is equivalent to a certain
topological constraint on the supergravity background that we shall discuss
later on. This is the case for D-branes which wrap supersymmetric cycles in all
Type II compactifications with vanishing cosmological constant \cite{BSspinc}.

\subsection{Anomalous Couplings}

The perturbative, chiral gauge anomaly on the brane-antibrane worldvolume may
be computed as the index of an appropriate Dirac operator \cite{AGDW}. For
this, we first need to identify the massless spectrum in the brane-antibrane
worldvolume field theory. As we have mentioned, after the GSO projection, there
remains chiral fermion zero modes in the $p$-$p$ and
$\overline{p}$-$\overline{p}$ open string sectors, but not in the
$p$-$\overline{p}$ sectors. In these latter sectors there are of course the
superpartners of the tachyon fields, which are massless bi-fundamental fermion
fields of opposite chirality. The fermions are all in a one-to-one
correspondence with the relevant open string Ramond sector ground states. Open
string quantization requires them to be sections of the spinor bundle lifted
from the spacetime tangent bundle $T(X)$ restricted to $\Sigma$,
$T(X)|_\Sigma=T\Sigma\oplus N\Sigma$, where $N\Sigma$ is the normal bundle to
$\Sigma$ in $X$. The GSO projection restricts the fermions to have definite and
opposite chiralities in the $p$-$p$ and $\overline{p}$-$\overline{p}$ sectors,
and also in the $p$-$\overline{p}$ and $\overline{p}$-$p$ sectors, with respect
to the local spacetime Lorentz group $SO(9,1)$. Upon dimensional reduction from
the spacetime manifold $X$ to the worldvolume manifold $\Sigma\subset X$, the
spacetime Lorentz group is broken to the $p+1$-dimensional local Lorentz group
of $\Sigma$ plus a global R-symmetry group corresponding to the structure group
of the normal bundle $N\Sigma$, i.e. $SO(9,1)\to SO(p,1)\times SO(9-p)$. This
implies that a section of $T(X)$ in a representation $R$ when restricted to
$\Sigma$ will decompose into sections of $T\Sigma\otimes N\Sigma$ in
representations $R_T^a\otimes R_N^a$. In particular, a chiral spinor field on
$X$ will decompose into a multiplet of chiral fermion fields transforming under
the adjoint representation
\beq
\rho=\left({\bf N}^+\otimes\overline{{\bf N}^+}\right)\oplus\left({\bf
N}^-\otimes\overline{{\bf N}^-}\right)\oplus\left({\bf
N}^+\otimes\overline{{\bf N}^-}\right)\oplus\left({\bf
N}^-\otimes\overline{{\bf N}^+}\right)
\label{rho}\eeq
of the brane-antibrane Chan-Paton structure group $U(N^+)\times U(N^-)$. This
is the structure that is required of the superpartners for the lowest
components of the superconnection gauge fields (\ref{calAmatrixgen}). The
Chan-Paton bundles $E^\pm$ are combined into the superbundle $E=E^+\oplus E^-$
and tensored with the appropriate spinor bundle. The above arguments imply that
the total $\zed_2$-grading is then the canonical tensor product grading, as in
(\ref{twistspingrad}), i.e. the chiral fermion fields on $\Sigma$ are the
sections
\beq
\Psi=\Psi_++\Psi_-~~~~~~,~~~~~~\Psi_\pm=\pmatrix{\psi^+_{(\mp)}\cr
\psi^-_{(\pm)}\cr}\in C^\infty(\Sigma,S_E^\pm) \ .
\label{spinorfields}\eeq

The fluctuations of these chiral fermion fields lead to quantum anomalies in
the brane-antibrane worldvolume effective field theory. There are several ways
to argue that these are the only anomalies produced. For instance, one may
argue as we have above from the basic properties of the GSO projection, or
alternatively by examining the massless spectrum of the intersection of two
brane-antibrane systems as we will briefly describe later on. One subtlety
concerns the fact that the brane-antibrane system is actually unstable. The
fact that one is not sitting in the true vacuum of the theory in making these
arguments makes them a little suspect. However, it is believed that
supersymmetry is not a necessary requirement in identifying these chiral
fermionic zero modes, i.e. even for non-supersymmetric brane configurations it
is still possible to correctly capture the massless fermionic content as above
\cite{SchwarzW,cy}. In any case, we assume that this is indeed the appropriate
ground state that remains after tachyon condensation on the brane-antibrane
system. This assumption may be motivated by the fact that it will lead to the
appropriate massless spectrum of induced lower dimensional D-brane charges that
remain after the tachyonic Higgs mechanism.

We may now proceed to use the standard topological index formula \cite{AS1}.
For this, we consider the appropriate spinor bundle ${\cal S}(\Sigma)={\cal
S}(T\Sigma)\otimes{\cal S}(N\Sigma)={\cal S}^+(\Sigma)\oplus{\cal
S}^-(\Sigma)$, where ${\cal S}(T\Sigma)$ and ${\cal S}(N\Sigma)$ are the spinor
bundles lifted from the tangent and normal bundles to $\Sigma$ in $X$,
respectively, and we use the standard tensor product grading
\beq
{\cal S}^\pm(\Sigma)=\left({\cal S}^\pm(T\Sigma)\otimes{\cal
S}^+(N\Sigma)\right)\oplus\left({\cal S}^\mp(T\Sigma)\otimes{\cal
S}^-(N\Sigma)\right) \ .
\label{calSpm}\eeq
The open string ground states in the Ramond sector are sections of ${\cal
S}(T\Sigma)$, while those of the Neveu-Schwarz sector are sections of ${\cal
S}(N\Sigma)$. We then incorporate the Chan-Paton superbundle by defining ${\cal
E}={\cal S}(\Sigma)\otimes E_\rho={\cal E}^+\oplus{\cal E}^-$, where $\rho$ is
the $U(N^+)\times U(N^-)$ representation (\ref{rho}) carried by the fermionic
open string zero modes. The appropriate grading is then taken to be
\beq
{\cal E}^\pm=\left({\cal S}^\pm(\Sigma)\otimes E_\rho^+\right)\oplus\left({\cal
S}^\mp(\Sigma)\otimes E_\rho^-\right) \ .
\label{calEpm}\eeq
The Dirac operator $\Dirac$, constructed as above from the appropriate lifted
spinor bundles, defines the two-term complex
\beq
 C^\infty(\Sigma,{\cal E}^\pm)~\stackrel{\Dirac}{\longrightarrow}~
C^\infty(\Sigma,{\cal E}^\mp) \ ,
\label{twotermcomplex}\eeq
and the standard index theorem applied to the superbundle ${\cal E}\to\Sigma$
yields
\beq
\ind\,i\Dirac=(-1)^{(p+1)(p+2)/2}\int\limits_\Sigma\ch^+({\cal
E})\wedge\frac{{\rm Td}(T\Sigma\otimes\complex)}{\chi(T\Sigma)} \ .
\label{indexthm}\eeq
Here $\ch^+({\cal E})$ denotes the Chern character of the superbundle $\cal E$
which may be represented by the closed differential form \cite{Quillen1}
\beq
\ch^+({\cal E})=\Tr^+\,\exp\frac1{2\pi i}\,F_\sSE \ ,
\label{chernplus}\eeq
where $\SE$ is a superconnection on $\cal E$ and the supertrace is taken over
the $\zed_2$-graded Hilbert space ${\cal H}={\cal H}^+\oplus{\cal H}^-$, with
${\cal H}^\pm= C^\infty(\Sigma,{\cal E}^\pm)$.\footnote{\baselineskip=12pt More
precisely, $\hil$ is the $L^2$-norm completion of the space of smooth sections
on $\Sigma\to{\cal E}$ with inner product defined with respect to the
Riemannian volume form $d{\rm vol}(\Sigma)$ of $\Sigma$.} The grading
automorphism (\ref{gradingop}) satisfies
$\varepsilon\Dirac=-\Dirac\varepsilon$. The second factor in (\ref{indexthm})
is a standard characteristic class which depends only on the topology of the
brane worldvolume manifold. For any oriented real vector bundle $V\to\Sigma$,
$\chi(V)$ denotes the cohomological Euler class of $V$, i.e. the restriction of
the Thom class $\Phi(V)$ to the zero section $H^{p+1}(V,\zed)\cong
H^{p+1}(\Sigma,\zed)$, while ${\rm Td}(V\otimes\complex)$ is the Todd class of
the complexification of $V$. The latter class may be related to
Atiyah-Hirzebruch class $\widehat{A}(V)$ by
\beq
\widehat{A}(V)=\e^{d(V)/2}\wedge\,\sqrt{{\rm Td}(V\otimes\complex)} \ ,
\label{ATdrel}\eeq
where the degree two integral characteristic class $d(V)$ determines a spin$^c$
structure on $V$ and may be defined as follows. The group homomorphism
\beq
Spin^c(p+1)=\Bigl(Spin(p+1)\times U(1)\Bigr)\,/\,\zed_2~\longrightarrow~U(1)
\label{Spinchomo}\eeq
induces a map $H^1(\Sigma,Spin^c(p+1))\to H^1(\Sigma,U(1))$ on cohomology, from
which we may associate a complex line bundle $L_V\to\Sigma$ to the vector
bundle $V$. Then $d(V)$ is defined as the first Chern class $c_1(L_V)$ of this
line bundle, and its mod 2 reduction yields the second Stiefel-Whitney class
$w_2(V)$.

The index formula (\ref{indexthm}) can be expanded by using the fact that the
Chern character respects the semi-ring structure as a map $V\mapsto\ch^+(V)$
from bundles to cohomology classes, i.e. $\ch^+(V\oplus W)=\ch^+(V)+\ch^+(W)$
and $\ch^+(V\otimes W)=\ch^+(V)\wedge\ch^+(W)$, so that
\beq
\ch^+({\cal E})=\ch_\rho^+(E)\wedge\ch^+\Bigl({\cal
S}(T\Sigma)\Bigr)\wedge\ch^+\Bigl({\cal S}(N\Sigma)\Bigr) \ ,
\label{chmult}\eeq
where $\ch_\rho^+$ denotes the Chern character as in (\ref{chernplus}) but with
the supertrace taken in the representation (\ref{rho}) of the Chan-Paton
structure group. Furthermore, if $V\to\Sigma$ is any oriented real spin$^c$
bundle and ${\cal S}(V)={\cal S}^+(V)\oplus{\cal S}^-(V)$ the spinor bundle
lifted from $V$, then we have the identity
\beq
\ch^+\Bigl({\cal S}(V)\Bigr)=\e^{d(V)}\wedge\frac{\chi(V)}{\widehat{A}(V)} \ .
\label{chSVid}\eeq
Combining the above relations, we may write (\ref{indexthm}) finally as
\beq
\ind\,i\Dirac=\int\limits_\Sigma\ch^+(E)\wedge\ch^+(\overline{E})\wedge
\e^{d(N\Sigma)}\wedge\frac{\widehat{A}(T\Sigma)}{\widehat{A}(N\Sigma)}
\wedge\chi(N\Sigma) \ ,
\label{indexfinal}\eeq
where $\overline{E}$ is the complex conjugate of the superbundle $E$. The Chern
characters in (\ref{indexfinal}) may be represented in terms of supertraces in
the fundamental representation of $U(N^+)\times U(N^-)$ and the curvature
$F_\sSA$ of a superconnection on $E$ \cite{Quillen1}, as described in the
previous section. We have used the properties
$\ch^+_{\rho_1\otimes\rho_2}(V)=\ch^+_{\rho_1}(V)\wedge\ch^+_{\rho_2}(V)$ and
$\ch_{\overline{\rho}}^+(V)=\ch^+_\rho(\overline{V})$ for unitary gauge bundles
$V\to\Sigma$. Note that the class $d(N\Sigma)$ also determines the spin$^c$
structure on $\Sigma$ itself \cite{WittenK,FW}.

The perturbative chiral gauge anomaly is related to the index
(\ref{indexfinal}) in the usual way \cite{AS3,AGG} by the descent formula
$A=2\pi i\,(\ind\,i\Dirac)^{(1)}$ where, for any invariant polynomial $I$ of
the Chan-Paton gauge bundle $E$, $I^{(1)}$ denotes its Wess-Zumino descendent
which is constructed as in section~2.4. Namely, we decompose $I=I_0+dI^{(0)}$
locally, where $I_0$ is the constant part of $I$ and $I^{(0)}$ is its secondary
characteristic. Then $I^{(1)}$ is defined via the gauge variation $\delta_{\rm
BRST}I^{(0)}=dI^{(1)}$. In fact, the derivation given above can be applied
straightforwardly to compute the anomaly on an ordinary D-brane system, by
noting that {\it any} vector bundle $V\to\Sigma$ can be trivially
$\zed_2$-graded by setting $V^+=V$ and $V^-=\Sigma\times\{0\}$. This defines an
even grading with grading automorphism $\varepsilon=\id$. The Chern characters
(\ref{chernplus}) then coincide with the usual (ungraded) ones. As is the usual
case, these quantum anomalies should cancel the classical anomalies which arise
due to the magnetic RR interactions of D-branes. This standard argument
\cite{MMK},\cite{GMM}--\cite{ss} is of a cohomological nature and can be
straightforwardly adapted to brane-antibrane systems~\cite{SchwarzW}. We will
therefore be brief.

Given a closed brane-antibrane worldvolume $\Sigma$, we postulate a coupling to
the spacetime Ramond-Ramond $p$-form fields $C_{(p)}$ of the form
(\ref{WZansatz}), where the coupling $\cal Y$ is determined by demanding that
the classical and quantum anomalies cancel each other. We integrate
(\ref{WZansatz}) by parts and rewrite it in terms of the constant part ${\cal
Y}_{0}$ and the descendent ${\cal Y}^{(0)}$ of ${\cal Y}$, i.e. ${\cal Y}={\cal
Y}_{0}+d{\cal Y}^{(0)}$ with $\delta_{\rm BRST}{\cal Y}^{(0)}=d{\cal Y}^{(1)}$.
The zero mode ${\cal Y}_{0}$ may be set to unity by suitably normalizing the
charge $\mu$. With $G=d\cal C$ the total RR field strength, we then have
\beq
Z=-\frac{\mu}2\,\int\limits_X\delta(\Sigma)\wedge\left({\cal C}-(-1)^\sigma
\,G\wedge{\cal Y}^{(0)}\right)
\label{WZiparts}\eeq
where $\sigma=1$ (resp. $\sigma=0$) for Type IIA (resp. Type IIB) D-branes, and
$\delta(\Sigma)$ is the deRham current which is a delta-function supported
representative of the Poincar\'e dual cohomology class to the embedding
$\Sigma\stackrel{\phi}{\hookrightarrow}X$. Globally, $\delta(\Sigma)$ is a
section of the normal bundle $N\Sigma$ with compact support. It may be
represented locally on $\Sigma$ by taking the zero section of $N\Sigma$. On the
other hand, in cohomology $\delta(\Sigma)$ may be identified with the Thom
class $\Phi(N\Sigma)$ of the normal bundle whose zero section is the Euler
class $\chi(N\Sigma)$. It follows that the deRham current possesses the global
property \cite{cy}
\beq
\delta(\Sigma)\wedge\delta(\Sigma)=\delta(\Sigma)\wedge\chi(N\Sigma) \ .
\label{deRhamglobal}\eeq

The coupling (\ref{WZiparts}) modifies the RR equations of motion and Bianchi
identity. In particular, $G$ is no longer a closed form, because
\beq
dG=-\mu\,\delta(\Sigma)\wedge\overline{{\cal Y}}
\label{Hbianchi}\eeq
where $\overline{{\cal Y}}$ is obtained from ${\cal Y}$ by complex conjugation
of the corresponding Chan-Paton gauge group representation. The minimal
expression for the field strength $G$ is then
\beq
G=d{\cal C}-(-1)^\sigma\,\mu\,\delta(\Sigma)\wedge\overline{{\cal Y}}^{(0)}
\label{HdCmin}\eeq
with $\overline{{\cal Y}}^{(0)}$ the secondary characteristic of
$\overline{{\cal Y}}$. By demanding that $G$ be gauge invariant, it follows
that the potential $\cal C$ must acquire an anomalous gauge transformation in
order to compensate the gauge variation of the second term in (\ref{HdCmin}),
\beq
\delta_{\rm BRST}{\cal C}=\mu\,\delta(\Sigma)\wedge\overline{{\cal Y}}^{(1)}
\label{anomvarC}\eeq
where $\overline{{\cal Y}}^{(1)}$ is the Wess-Zumino descendent of
$\overline{{\cal Y}}$. Thus, under a gauge transformation $\delta_{\rm BRST}$,
one finds that the RR couplings (\ref{WZiparts}) yield a gauge anomaly given by
\beq
\delta_{\rm BRST}Z=-\frac{\mu^2}2\,\int\limits_X\delta(\Sigma)\wedge
\delta(\Sigma)\wedge\left({\cal Y}\wedge\overline{{\cal Y}}\right)^{(1)} \ .
\label{gaugeanomS}\eeq
By using (\ref{deRhamglobal}) we find that the magnetic RR coupling on $\Sigma$
is anomalous.

The corresponding classical anomaly inflow is given by $A=2\pi
i\int_{\Sigma}I^{(1)}$, where
\beq
I=-\frac{\mu^2}{4\pi}\,{\cal Y}\wedge\overline{{\cal Y}}\wedge\chi(N\Sigma) \ .
\label{Ikl}\eeq
The anomolous form (\ref{Ikl}) is of the same type as the integrand of
(\ref{indexfinal}), and it implies that the anomalous RR coupling on the
brane-antibrane system is given by
\beq
{\cal Y}=\ch^+(E)\wedge\e^{d(N\Sigma)/2}\wedge
\sqrt{\frac{\widehat{A}(T\Sigma)}{\widehat{A}(N\Sigma)}} \ .
\label{calYfinal}\eeq
The classical and quantum anomalies will thereby cancel provided that
$\mu=\sqrt{2\pi}$, which is the correct quantum of RR charge (in string units).
Furthermore, these choices cancel the anomalies which arise from the magnetic
RR interactions between two different brane-antibrane systems, i.e. on the
I-branes $\Sigma_{12}=\Sigma_1\cap\Sigma_2$~\cite{MMK},\cite{GMM}--\cite{ss}.
Then, in addition to the chiral massless fermions required by open string
quantization and the GSO projection, there are massless spinors which arise
from the open string sectors which start in $\Sigma_1$ and end in $\Sigma_2$.
Because of the previously stated properties of the GSO projection, the only
fermionic zero modes which can arise in this way come from the open string
sectors which begin on a brane (resp. antibrane) and end on another brane
(resp. antibrane). Thus the index theoretical calculation on an I-brane
$\Sigma_{12}$ will go through in exactly the same way as before, and produce a
quantum anomaly which cancels the classical I-brane anomaly inflow as above.

Although the derivation above formally captures the dependence of the RR
couplings on the superconnection gauge fields, it should be noted that while
they do naturally arise from the pertinent index theorem, the anomaly inflow
arguments are not sufficient to determine the tachyon terms in the
action.\footnote{\baselineskip=12pt The author is grateful to F. Larsen for
pointing this out to him.} This follows from the fact that inflow determines
the couplings only up to gauge invariant terms. One can change representative
of the cohomology class (\ref{calYfinal}) to ${\cal Y}+d{\cal V}$, where $\cal
V$ is any local, gauge-invariant polynomial. For closed Ramond-Ramond
potentials $\cal C$, such extra terms do not contribute to the charge. Since
the tachyon dependent terms in the superconnection Chern character
(\ref{chernplus}) are precisely of this gauge-invariant form, the tachyon
dependent parts of the RR-couplings can be cancelled in this way. Of course,
this immediately follows from the fact, proven in section~2.4, that the
cohomology classes generated by $\ch^+(E)$ are independent of the tachyon field
$T$. Put differently, the worldvolume fermionic zero-modes are not charged with
respect to the tachyon background. Nevertheless, since the usual anomaly
analysis, based on the index theorem, applied to the worldvolume gauge fields
involves the Chern character of a superbundle, we shall choose the most general
representative which involves a non-zero tachyon field $T$. As we will see,
this formula leads to the correct results for the tachyonic couplings on
non-BPS systems of branes and, moreover, determines the appropriate
relationship between D-brane charges and K-homology.

\subsection{Complete Chern-Simons Action}

Before writing down the final version of the anomalous coupling on a
brane-antibrane system, there are some aspects of the above derivation that we
should first discuss. We shall work in the static gauge of the worldvolume
diffeomorphism group which may be defined as follows. We split the local
coordinates of $X$ into longitudinal and transverse coordinates with respect to
$\Sigma$, $x^a=(x^\mu,x^i)$, and use spacetime diffeomorphism invariance to fix
$\Sigma$ at the coordinates $x^i=0$ for $i=p+1,\dots,9$. Then we use
worldvolume diffeomorphism invariance to identify the longitudinal coordinates
with those of $\Sigma$, $x^\mu=\xi^\mu$ for $\mu=0,1,\dots,p$. For multiple
branes and antibranes, we should identify the transverse coordinates to the
worldvolume $\Sigma$ with matrices in the Lie algebra of the brane-antibrane
gauge group $U(N^+)\times U(N^-)$, $x^i={\cal X}^i$, where
\beq
{\cal X}^i=\pmatrix{\phi_+^i&0\cr0&\phi_-^i\cr} \ .
\label{calXdef}\eeq
The $N^+\times N^+$ (resp. $N^-\times N^-$) Hermitian matrices $\phi_+^i$
(resp. $\phi_-^i$) describe the $9-p$ transverse degrees of freedom of the
branes (resp. antibranes). They transform in the adjoint representation of
$U(N^+)$ (resp. $U(N^-)$) and correspond to the fields of the R-symmetry group
$SO(9-p)$ of the dimensionally reduced Yang-Mills gauge theory to the
brane-antibrane worldvolume. From a global perspective, we may use the
Riemannian structure on the spacetime manifold $X$ to identify the normal
bundle $N\Sigma$ with a tubular neighbourhood of the worldvolume $\Sigma$ in
$X$. Then the transverse degrees of freedom of the brane-antibrane system
wrapping $\Sigma$ which are described by sections of $N\Sigma$ are augmented to
sections of the $U(N^+)\times U(N^-)$ bundle $N\Sigma\otimes({\rm
End}(E^+)\oplus{\rm End}(E^-))$. Note that the total brane-antibrane scalar
fields (\ref{calXdef}) are block-diagonal because the GSO projection eliminates
the scalar degrees of freedom in the $p$-$\overline{p}$ open string sectors. In
these sectors there are of course the remnant tachyon fields, but these are
objects which live in the worldvolume theory itself and are not attributed to
the transverse degrees of freedom of the branes and antibranes. We shall see
below how to include the modifications due to the tachyonic degrees of freedom.

The standard definition of pull-backs should then be altered so as to replace
all transverse coordinates with the matrices ${\cal X}^i$ and all worldvolume
derivatives with covariant ones \cite{myers,hassan}. In order to obtain a
covariant expression, we must also account for the possible non-trivial normal
bundle topology and the fact that the transverse scalar fields are really
sections of $N\Sigma$. Let $\alpha_i^m$, $m=p+1,\dots,9$, span a frame in
$N\Sigma$, and introduce the connection $\Theta_{\mu
n}^m=\alpha_j^m\,\partial_\mu\alpha_n^j$ on the normal bundle to the
brane-antibrane worldvolume. Then, on the Ramond-Ramond fields, the definition
of the pull-back $\phi^*$ induced by the embedding
$\Sigma\stackrel{\phi}{\hookrightarrow}X$ is taken to be
\beq
\left(\phi^*\,C_{(p)}\right)_{\mu_1\cdots\mu_p}=C_{(p)\mu_1\cdots\mu_p}+
\sum_{k=1}^pC_{(p)i_1\cdots i_k\{\mu_1\cdots\mu_{p-k}}\,\nabla^{\rm N}
_{\mu_{p-k+1}}{\cal X}^{m_1}\,\alpha_{m_1}^{i_1}\cdots
\nabla^{\rm N}_{\mu_p\}}{\cal X}^{m_k}\,\alpha_{m_k}^{i_k} \ ,
\label{pullbacks}\eeq
where
\beq
\nabla^{\rm N}_\mu{\cal X}^m\,\alpha_m^i=
\left(D_\mu{\cal X}^m+\Theta_{\mu n}^m\,{\cal X}^n\right)\alpha_m^i
\label{nablaNdef}\eeq
and $D_\mu$ is the covariant derivative defined as in (\ref{DTdef}). This
produces a non-trivial interaction between the Ramond-Ramond fields and the
non-abelian transverse excitations of the branes and antibranes. In the case of
multiple D-branes alone, there are in addition multipole terms and other
commutator terms which couple to the background supergravity fields
\cite{myers}. These terms are required by T-duality and in order to match
results from Matrix theory. In the present case, we will impose such
requirements, in addition to the pull-back definition (\ref{pullbacks}), for
the sake of matching with the recent observations concerning D-brane effective
actions. The matrix structure of the transverse coordinates for multiple branes
and antibranes will become important later on and will lead to a D-brane action
which is explicitly T-duality invariant.

First of all, the background spacetime fields restricted to the worldvolume
$\Sigma$ are formally regarded as functions of the transverse coordinates,
under the identification $x^i={\cal X}^i$. This is achieved by using the formal
Taylor series expansions of the fields in the transverse coordinates and it
defines couplings of their multipole moments to the adjoint scalar fields
${\cal X}^i$. For instance, for the RR tensor potentials we write
\beq
C_{(p)a_1\dots a_p}=C_{(p)a_1\dots a_p}(\xi^\mu,0)+\sum_{n=1}^\infty
\frac1{n!}\,\frac\partial{\partial x^{i_1}}\cdots\frac\partial{\partial
x^{i_n}}\,C_{(p)a_1\cdots a_p}(\xi^\mu,x^i)\biggm|_{x^i=0}\,{\cal
X}^{i_1}\cdots{\cal X}^{i_n} \ .
\label{Cpexpand}\eeq
The couplings in (\ref{Cpexpand}) may be obtained by application of the
operator $\exp\jmath^{~}_{\cal X} d_\perp|_{x^\perp=0}=\exp{\cal
X}^i\frac\partial{\partial x^i}|_{x^i=0}$ to the field $C_{(p)}$, where
$\jmath^{~}_{\cal X}$ is the interior multiplication operator with respect to
${\cal X}^i$ of degree $-1$ and $d_\perp$ the exterior derivative on the normal
bundle $N\Sigma$. Of course, in the case $N^+=N^-=1$ the scalar fields
$\phi_+^i,\phi_-^i$ coincide with the transverse coordinates $x^i$ themselves
and all higher partial wave couplings in (\ref{Cpexpand}) disappear. In
addition, we need to insert a coupling of the background fields to commutators
of the scalar fields (\ref{calXdef}). This is again achieved via action of the
interior multiplication operator as $\e^{i(\jmath^{~}_{\cal X})^2}C_{(p)}$.
Using antisymmetry of the components of the differential form $C_{(p)}$, the
$n$-th term (with $2n\leq p$) in the expansion of this object is given by
\beq
\left[(\jmath^{~}_{\cal X})^{2n}\,C_{(p)}\right]_{a_1\cdots a_{p-2n}}=
\frac1{2^n{p\choose2n}}\,C_{(p)i_1\cdots i_{2n}a_1\cdots a_{p-2n}}\,
\Bigl[{\cal X}^{i_1}\,,\,{\cal X}^{i_2}\Bigr]\cdots\Bigl[{\cal
X}^{i_{2n-1}}\,,\,{\cal X}^{i_{2n}}\Bigr] \ .
\label{jXCpdef}\eeq
In addition, T-duality invariance requires the background fields to couple to
the tachyon field, because such couplings are induced by T-duality
transformations of the one-form parts ${\cal F}_{(1)}$ of the superconnection
field strengths in (\ref{covderivT}) \cite{TTU,myersnon}. The appropriate
modification comes from replacing the operator $(\jmath^{~}_{\cal X})^2$ by
\beq
{\cal J}_{\cal X}(T)^2=(\jmath^{~}_{\cal X})^2+i\left[\jmath^{~}_{\cal X}~,~
\pmatrix{0&T^\dagger\cr T&0\cr}\right] \ .
\label{calJcalXT}\eeq

Having described the appropriate physical alterations of the anomalous
couplings which must be made for multiple branes and antibranes, we now turn to
a discussion of the factor $\e^{d(N\Sigma)/2}$ in (\ref{calYfinal}), which
accounts for the spin$^c$ structure on $\Sigma$ and can induce charge shifts of
degree two on the brane-antibrane worldvolume $\Sigma$. If $\Sigma$ is a
connected almost complex manifold, then the class $d(N\Sigma)\in
H^2(\Sigma,\zed)$ can be represented by the first Chern class of the normal
bundle as $d(N\Sigma)=-c_1(N\Sigma)$. If the worldvolume $\Sigma$ were not a
spin$^c$ manifold, then one would have to incorporate a topologically
non-trivial Neveu-Schwarz two-form field $B$ into the string background in
order to cancel certain worldsheet anomalies \cite{WittenK,WittenAdS,FW}. The
result of this cancellation is that it trivializes a certain line bundle over
the loop space of the brane worldvolume which is defined by two-cycle
holonomies of $B$. In the purely bosonic case this would imply that the
$B$-field restricted to $\Sigma$ is necessarily topologically trivial
\cite{FW}. For the full superstring theory, this is not the case, but the
topological type of $B$ restricted to $\Sigma$ is uniquely determined by the
bundle trivialization property. A $B$-field on $\Sigma$ is classified
topologically by its characteristic class $\phi^*[H]\in H^3(\Sigma,\zed)$ which
as a differential form is represented by the pull-back of the field strength
$H=dB$. $B$ being topologically trivial means that $\phi^*[H]$ vanishes as an
integral cohomology class, and not only in real cohomology. The bundle
trivialization just mentioned is equivalent to the condition
$\phi^*[H]=W_3(\Sigma)$, where $W_3(\Sigma)\in H^3(\Sigma,\zed)$ is the
Dixmier-Douady invariant which may be defined as the image of the second
Stiefel-Whitney class $w_2(\Sigma)\in H^2(\Sigma,\zed_2)$ under the appropriate
connecting homomorphism (the Bockstein map) in cohomology
\cite{WittenK,FW,WittenAdS}. The cancellation of global worldvolume anomalies
in the previous subsection required that $\Sigma$ admit a spin$^c$-structure,
which is equivalent to $W_3(\Sigma)=0$. Thus the restriction of $B$ to $\Sigma$
is topologically trivial. Furthermore, in that case, in order to cancel the
worldsheet anomalies requires that the brane worldvolume gauge fields define
spin$^c$ connections, rather than single-valued ones. But this is precisely
what we assumed in section~2 when we used the superconnection gauge field $\cal
A$ to define a Clifford superconnection $\SS$, and ultimately the appropriate
Dirac operator $\Sdirac$ which allowed us to compute the anomalous couplings to
the RR fields. These remarks illustrate the consistency of the present analysis
thus far. It would be interesting to extend the analysis to topologically
non-trivial $B$-fields and hence worldvolumes $\Sigma$ which do not admit
spin$^c$-structures.

For the purposes of analysing the anomalies in the brane-antibrane worldvolume
field theories, we should therefore incorporate a topologically trivial
Neveu-Schwarz two-form field. This is included via two-cycle holonomies of $B$.
The final object we need to take special care of is the topological normal
bundle correction term in (\ref{calYfinal}). This can be properly incorporated
by covariantizing the couplings that we have described above \cite{hassan}.
Defining ${\cal C}'={\cal C}\wedge\e^{-\phi^*B/2\pi i}$, taking into account
the non-abelian dynamics of D-branes amounts (for topologically trivial NS-NS
$B$-field) to replacing in the action (\ref{WZansatz}) the exterior product
${\cal C}'\wedge{\cal Y}$ by Clifford multiplication defined through the symbol
map
\beq
\sigma_{{\cal C}'}({\cal Y})={\cal C}'\wedge{\cal Y}-\jmath^{~}_{{\cal C}'}
{\cal Y} \ ,
\label{Cliffprod}\eeq
when the couplings are expressed in terms of bulk quantities as in
(\ref{SWZKtheory}). This overall modification by the spin geometry of $X$ fits
very nicely into the present formalism. However, we will not write the Clifford
multiplication explicitly and only assume its presence implicitly when we write
exterior products with ${\cal C}'$.

We have thus found that the anomalous, Chern-Simons couplings on a
brane-antibrane system wrapping a worldvolume $\Sigma$ of dimension $p+1$ are
given by
\bea
Z&=&-\frac{\sqrt{2\pi}}2\,\int\limits_\Sigma\Tr\,(-1)^F~\phi^*
\left(\exp\left\{i\,(\jmath^{~}_{\cal X})^2-\left[\jmath^{~}_{\cal X}~,~
\pmatrix{0&T^\dagger\cr T&0\cr}\right]\right\}~\e^{\jmath^{~}_{\cal X}d_\perp}
\biggm|_{x^\perp=0}\right.\nn\\&&\times\left.
\sum_{p=0}^{4+\sigma}C_{(2p+1-\sigma)}\wedge
\e^{-B/2\pi i}\right)\wedge\exp\frac1{2\pi i}\,F_\sSA\wedge\,
\sqrt{\widehat{A}(R_T)/\widehat{A}(R_N)}\wedge\e^{d(N\Sigma)/2} \ . \nn\\&&
\label{WZfinal}\eea
The trace in (\ref{WZfinal}) is taken in the fundamental representation of the
$U(N^+)\times U(N^-)$ gauge group, with $(-1)^F$ the grading automorphism of
the brane-antibrane pairs and $F_\sSA$ the field strength (\ref{FAred}) of the
superconnection which depends on the worldvolume gauge fields and the
brane-antibrane tachyon field (Here we take ${\cal G}=\id$ in
(\ref{Diracmatrix})). The matrix products in the argument of the trace in
(\ref{WZfinal}) must be given an appropriate ordering prescription, which we
take to be the symmetrized trace defined by (\ref{symtrace}) \cite{myers}. This
trace symmetrization will also be implicitly assumed in the following. The
third term in (\ref{WZfinal}) gives the appropriate gravitational couplings of
the fields, with $R_T$ (resp. $R_N$) the Riemann curvature two-form of the
tangent (resp. normal) bundle, and
\beq
\widehat{A}(R)=\prod_{a\geq1}\frac{r_a}{\sinh r_a}
\label{AhatRexpl}\eeq
where $4\pi r_a$ are the skew-eigenvalues of $R_{ab}$. The action
(\ref{WZfinal}) agrees with the form of the brane-antibrane coupling originally
proposed in \cite{kw}. Notice, however, that the formula (\ref{WZfinal}) is
ambiguous with respect to on-shell terms such as the equation of motion for the
tachyon field. The superconnection formalism employed above gives a unique
off-shell prescription for the RR charges, alternatively to the boundary string
field theory prescription used in~\cite{KLDbarD}. The physical implications of
this will be that the index theoretical calculations automatically lead to the
relationships between D-brane charges and K-theory, as will be extensively
described in section~6.

\subsection{Tachyon Condensation}

There is another natural invariant action that can be constructed using the
superconnection formalism. For this, we consider the natural inner product
density $(\cdot,\cdot)$ on the algebra $\Omega(\Sigma,{\rm End}\,E)$ of
sections of the endomorphism bundle defined by
\beq
(A,B)=\Tr\left(B^\dagger\wedge*A\right) \ ,
\label{innprod}\eeq
where $*$ denotes the Hodge star-operator on $\Sigma$. This is the inner
product density that is canonically inherited from $\Omega(\Sigma)$. If
$\|\cdot\|$ denotes the corresponding norm density, then we can write down a
Euclidean action in the form
\bea
Z_{\rm kin}&=&\frac12\,\int\limits_\Sigma\left\|F_\sSA-{\cal G}
\right\|^2\nonumber\\
&=&\frac12\,\int\limits_\Sigma\left(\sum_{k\geq1}\left\|{\cal F}_{(k)}
\right\|^2+\left\|{\cal F}_{(0)}-{\cal G}\right\|^2\right) \ ,
\label{Skindef}\eea
where
\beq
{\cal G}=\pmatrix{{\cal G}^+&0\cr0&{\cal G}^-\cr}
\label{Gdef}\eeq
is a constant abelian flux with $({\cal G}^\pm)^\dagger={\cal G}^\pm$.
Expanding (\ref{Skindef}) out using (\ref{TTdagger})--(\ref{ordcurvs}) then
leads to
\bea
Z_{\rm kin}&=&\int\limits_\Sigma d{\rm vol}(\Sigma)~\Tr\left[\frac12\,
\left(F_{\mu\nu}^+\right)^2+\frac12\,\Bigl(F^-_{\mu\nu}\Bigr)^2+
\left|D_\mu T\right|^2\right.\nn\\& &+\left.
\frac12\,\left(T^\dagger T-{\cal G}^+\right)^2
+\frac12\,\left(TT^\dagger-{\cal G}^-\right)^2+\dots\right] \ .
\label{Euclaction}\eea
This is the general form of the brane-antibrane worldvolume action anticipated
from two-loop order, on-shell string theory scattering amplitudes
\cite{pesando}. Therefore, we see that the superconnection formalism also gives
a compact way of representing the kinetic terms in the low energy effective
field theory. Terms involving the worldvolume scalar fields (\ref{calXdef}) may
be incorporated in a manner analogous to that described in the previous
subsection~\cite{myers}. Higher order corrections to (\ref{Euclaction})
presumably come from a Born-Infeld expansion in powers $(F_\sSA)^n$ of the
superconnection curvature~\cite{AIO,Sen5,kluson2}.

In particular, from the second line of (\ref{Euclaction}) we obtain an explicit
expression for the tachyon potential. The minima of the action (\ref{Skindef})
determine the tachyon condensates $T_c$ which are given by the equation
\beq
F_\sSA={\cal G} \ .
\label{Tceq}\eeq
This condition requires, among other things, covariantly flat gauge field
configurations on the branes and antibranes, and a covariantly constant tachyon
field. There are two special cases whereby the variational equation
(\ref{Tceq}) can be solved with ease. If ${\cal G}=0$ there is a unique
solution $T_c=0$ giving a $U(N^+)\times U(N^-)$ invariant vacuum. In this case
there is no symmetry breaking. If ${\cal G}=m^2\,\id$ with $m^2>0$ and
$N^+=N^-=N$, then $T_c^\dagger\,T_c=T_c\,T_c^\dagger=m^2\,\id$ and $T_c$
establishes an isomorphism between the fiber spaces $E_x^+$ and $E_x^-$ of the
Chan-Paton superbundle $E\to\Sigma$. Conversely, any such isomorphism yields a
tachyon condensate $T_c$. This configuration breaks the $U(N)\times U(N)$ gauge
symmetry group down to its diagonal subgroup $U(N)_{\rm diag}$. The constant
operator $\cal G$ is a constant abelian flux on the brane-antibrane
worldvolume, and the tachyon field is asymptotically a bundle isomorphism
between the branes and antibranes where it reaches its vacuum expectation
value. Thus we recover the standard requirements for tachyon condensation in
brane-antibrane systems \cite{NonBPSrev,Sen1,WittenK,osrev,Sen2}. Here we have
derived them from a purely geometric formalism, which also allows for more
general symmetry breaking patterns.

The mechanism for symmetry breaking here is even more elementary, because it in
fact originates from the generalized Dirac operator
(\ref{Diracmaps})--(\ref{DiracQ}). To see this, we use the Lichnerowicz formula
for the ordinary Dirac operators (\ref{DiracQ}) to compute
\bea
\Sdirac^2&=&\left(\Delta_\Sigma+\frac14\,r_\Sigma\right)\,\id_{N^+|N^-}
\nn\\&&+\,
\pmatrix{Q\left(F^++T^\dagger T\right)&\dirac_A^+\left({\cal G}_{\rm s}
\otimes T^\dagger\right)+\left({\cal G}_{\rm s}
\otimes T^\dagger\right)\dirac_A^-\cr\dirac_A^-\Bigl({\cal G}_{\rm s}
\otimes T\Bigr)+\Bigl({\cal G}_{\rm s}
\otimes T\Bigr)\dirac_A^+&Q\left(F^-+TT^\dagger\right)\cr} \ ,
\label{Lich}\eea
where $\Delta_\Sigma$ is the Laplace-Beltrami operator and $r_\Sigma$ the
scalar curvature of the worldvolume $\Sigma$, and $Q=c\circ\sigma^{-1}$ is the
quantization map. From (\ref{Lich}) it follows that the Lagrangian of
(\ref{Skindef}) may be computed from the symbol map as
\beq
\left\|F_\sSA\right\|^2=\Tr\left[\sigma\circ c^{-1}(\Sdirac^2)\right]^2 \ .
\label{Fsymbol}\eeq
Therefore, both the topological and the Born-Infeld type action on the
brane-antibrane system can be derived from the Dirac operator $\Sdirac$. This
is not surprising, since as we mentioned in section~2.3 the Dirac operator
carries the same amount of information as its corresponding superconnection
\cite{BGV}. This property is even more apparent if we supersymmetrize the
action (\ref{Skindef}) by adding a fermion coupling
$\overline{\Psi}\,\Sdirac\,\Psi$, with $\Psi$ the spinor fields
(\ref{spinorfields}). Then the fermion masses are induced by the Dirac operator
and correspond to the tachyonic expectation values of the brane-antibrane
pairs. The mass matrix ${\cal G}_{\rm s}\otimes T_c$ of the fermion fields
originate from the quantum field theory of the $p$-$\overline{p}$ open string
ground states which are given by the operator (\ref{Diracmatrix}) \cite{hori}.
Therefore, all of the standard properties of tachyon condensation come from a
spectral action involving the relevant generalized Dirac operator. These facts
will be instrumental in the K-theory interpretation that we shall give in
section~6. Notice, however, that (\ref{Euclaction}) is {\it not} the full form
of the kinetic part of the worldvolume action. In particular, it only agrees
with the results of~\cite{KLDbarD} up to terms involving off-shell and also
field-redefinition ambiguities. The assumption that the kinetic terms can be
written in terms of superconnection (or, equivalently, generalized Dirac
operator) quantities alone implicitly imposes an on-shell requirement.

Having established that the unstable brane-antibrane system will decay via a
Higgs mechanism, let us now examine the corresponding reduction of the
Chern-Simons action (\ref{WZfinal}). First of all, we note that if the tachyon
field is absent, $T\equiv0$, then the action (\ref{WZfinal}) is a sum
$Z=Z_++Z_-$, where
\bea
Z_\pm&=&\mp\,\frac{\sqrt{2\pi}}2\,\int\limits_\Sigma\Tr_\pm\,\phi_\pm^*
\left(\e^{i(\jmath^{~}_{\phi_\pm})^2}~\e^{\jmath^{~}_{\phi_\pm}
d_\perp}\biggm|_{x^\perp=0}\,
\sum_{p=0}^{4+\sigma}C_{(2p+1-\sigma)}\wedge\e^{-B/2\pi i}\right)\wedge
\exp\frac1{2\pi i}\,F^\pm\nn\\& &\wedge\,
\sqrt{\widehat{A}(R_T)/\widehat{A}(R_N)}\wedge\e^{d(N\Sigma)/2} \ .
\label{WZsum}\eea
The $\pm$ indices label the contributions from the branes and antibranes,
respectively, so that $\Tr_\pm$ denotes the (symmetrized) trace in the
fundamental representation of $U(N^\pm)$. Thus when the branes are
well-separated from the antibranes (so that there are no massless open string
$p$-$\overline{p}$ modes), the total Ramond-Ramond charge is given as the sum
of RR charges on the branes and antibranes. This includes the extra multipole
couplings for non-abelian systems as is required by T-duality \cite{myers}.
Similarly, in this case the action (\ref{Euclaction}) decomposes into a sum of
Yang-Mills actions for the field strengths $F^{\pm}$ on the branes and
antibranes. An interesting feature to examine in this context is the critical
value of the tachyon field at which the open string $p$-$\overline{p}$ modes
become relevant again and $Z\neq Z_++Z_-$ \cite{BanksSuss}. Within the present
framework this is a difficult question to answer, however, because one would
need to use distinguishable D-branes with distinct worldvolume manifolds for
the branes and antibranes \cite{pesando}.

Now let us reinstate the tachyonic coupling and see how to realize a
$(p-2k)$-brane in the worldvolume $\Sigma$ of the $p$-$\overline{p}$ pairs
\cite{WittenK,osrev}. For this, we assume that all length scales of the problem
are much larger than the string scale. We set $N^+=N^-$ and let $\tilde\Sigma$
be a spin$^c$ submanifold of codimension $2k$ in $\Sigma$. Then the normal
bundle $N(\tilde\Sigma,\Sigma)$ to $\tilde\Sigma$ in $\Sigma$ has structure
group $SO(2k)$. Let $\tilde\Sigma'$ be a tubular neighbourhood of
$\tilde\Sigma$ in $\Sigma$. Since $\tilde\Sigma$ and $\Sigma$ only have
spin$^c$ structures defined on them, the spinor bundles ${\cal
S}^\pm(N(\tilde\Sigma,\Sigma))$ cannot in general be constructed globally. Let
$ L_N\to\tilde\Sigma$ be the complex line bundle corresponding to the integral
cohomology class $d(N(\tilde\Sigma,\Sigma))$, i.e. $c_1(
L_N)=d(N(\tilde\Sigma,\Sigma))$. Then, generally, the square root $ L_N^{1/2}$
(with $ L_N^{1/2}\otimes L_N^{1/2}= L_N$) also cannot be constructed globally.
When there is two-torsion in the cohomology group $H^2(\tilde\Sigma,{\bf Z})$,
there are different square roots of $L_N$ and hence more than one spin$^c$
structure for a given class $d(N(\tilde\Sigma,\Sigma))$. However, the twisted
spinor bundles
\beq
{\cal S}_{ L_N^{1/2}}^\pm= L_N^{1/2}\otimes{\cal S}^\pm\left[N\left(
\tilde\Sigma\,,\,\Sigma\right)\right]
\label{calScalN}\eeq
{\it do} exist as vector bundles over $\tilde\Sigma'$. This is the precise
meaning of the existence of a spin$^c$ structure for $\tilde\Sigma$, and also
of the vanishing of the global worldsheet anomaly for topologically trivial
$B$-field \cite{FW}. We recall once again from section~2 that the natural
geometrical objects on the bundle (\ref{calScalN}) are Clifford
superconnections, or equivalently generalized Dirac operators.

Let ${\cal L}\to\tilde\Sigma$ be a given complex line bundle. We extend $\cal
L$ over all of $\Sigma$, if necessary by using Swan's theorem to choose a
bundle $I\to\tilde\Sigma$ such that ${\cal L}\oplus I$ is trivial. Similarly,
if necessary, we choose a bundle ${\cal I}\to\tilde\Sigma$ such that ${\cal
S}_{ L_N^{1/2}}^-\oplus{\cal I}$ is trivial. Then both ${\cal L}\oplus I$ and
${\cal S}_{ L_N^{1/2}}^-\oplus{\cal I}$ are extendable as vector bundles to the
whole of $\Sigma$. For the Chan-Paton superbundle $E=E^+\oplus E^-$ over
$\Sigma$ we may then take
\beq
E^\pm={\cal L}\otimes{\cal S}^\pm_{ L_N^{1/2}}\oplus I\oplus{\cal I} \ .
\label{EcalLcalS}\eeq
To construct a tachyon field, we consider the generators $\Gamma_i$,
$i=1,\dots,2k$, of the complex Clifford algebra $\cliff_{2k}$ of the transverse
structure group, which satisfy the Euclidean Dirac algebra
\beq
\Gamma_i\Gamma_j+\Gamma_j\Gamma_i=2\delta_{ij}
\label{Gammaialg}\eeq
and which may be decomposed as
\beq
\Gamma_i=\pmatrix{0&\gamma_i^\dagger\cr\gamma_i&0\cr}
\label{Gammaidecomp}\eeq
with respect to the chirality ${\bf Z}_2$-grading of the spinor bundle. They
are viewed as elements of the unitary group $U(2^k)$. The tachyon field is then
the section of $E$ which is defined locally by Clifford multiplication as
\beq
T(x)=\id_{\cal L}\otimes\left(2\pi f(x)\,\sum_{i=1}^{2k}\gamma_i\,x^i\right)
\oplus\id_{I\oplus{\cal I}}
\label{ABStachyon}\eeq
for $x\in\tilde\Sigma'$, where $f(x)$ is a real-valued convergence factor which
is constant near $\tilde\Sigma$ (where $x^i=0$) and which behaves as
$f_\infty/\|x\|$, $f_\infty={\rm const.}$, near $\partial\tilde\Sigma'$ (where
$x^i\to\infty$). We then pick gauge connections $A^\pm$ on $E^\pm$ which
satisfy the finite energy conditions $DT=DT^\dagger=0$ near
$\partial\tilde\Sigma'$. These choices are the standard assumptions used for
tachyon condensation and the bound state construction of D-branes
\cite{NonBPSrev,Sen1,WittenK,osrev,Sen2}. Note that by substituting the profile
(\ref{ABStachyon}) into (\ref{TTdagger}) and using the Clifford relations
(\ref{Gammaialg},\ref{Gammaidecomp}), the zero-form component of the
supercurvature may be computed to be
\beq
{\cal F}_{(0)}(x)=\Bigl(2\pi f(x)\Bigr)^2\,\Gamma_i\Gamma_j\,
x^ix^j=\Bigl(2\pi f(x)\Bigr)^2\,\|x\|^2\,\id \ .
\label{calF0ABS}\eeq
The fact that (\ref{calF0ABS}) is proportional to the identity matrix in
Chan-Paton space is related to the fact that the brane-antibrane pairs with the
tachyon configuration (\ref{ABStachyon}) condense to a single brane of
codimension $2k$. On $\partial\tilde\Sigma'$, where the tachyon field
(\ref{ABStachyon}) assumes its vacuum expectation value, the Higgs mass may
thereby be determined explicitly to be $m^2=(2\pi f_\infty)^2$.

The resulting configuration represents the desired superconnection $\SA$ to be
used in the topological action (\ref{WZfinal}), where we now neglect the
non-abelian transverse corrections (and hence generically ruin T-duality
invariance). The gravitational couplings may be simplified by using
multiplicativity of the characteristic classes, along with the Whitney sum
decompositions $T\Sigma|_{\tilde\Sigma}=T\tilde\Sigma\oplus
N(\tilde\Sigma,\Sigma)$ and $N\tilde\Sigma=N(\tilde\Sigma,\Sigma)\oplus
N\Sigma$. The Chern character may be simplified by using multiplicativity, the
definition (\ref{EcalLcalS}), and (\ref{calF0ABS}) to compute the
superconnection field strength (\ref{FAred}) with the finite energy conditions.
It is then straightforward to arrive at
\bea
Z&=&-\frac{\sqrt{2\pi}}2\,\int\limits_{\tilde\Sigma}~\sum_{p=0}^{4+\sigma}
\phi^*\left(C_{(2p+1-\sigma)}\wedge\e^{-B/2\pi i}\right)\wedge\ch({\cal L})
\wedge\sqrt{\frac{\widehat{A}\left(T\tilde\Sigma\right)}
{\widehat{A}\left(N\tilde\Sigma\right)}}\wedge\e^{d\bigl(N\tilde\Sigma
\bigr)/2}\nonumber\\&&\times\,\int\limits_{\tilde\Sigma'}
\e^{-2\pi f(x)^2\|x\|^2}~
\widehat{A}\left[N\left(\tilde\Sigma\,,\,\Sigma\right)\right]\wedge\e^{-c_1(
 L_N)/2}\wedge\left[\ch\left({\cal S}_{ L_N^{1/2}}^+\right)-
\ch\left({\cal S}_{ L_N^{1/2}}^-\right)\right] \ , \nn\\&&
\label{WZred}\eea
where in this formula ch is the ordinary (ungraded) Chern character. The second
integral in (\ref{WZred}) can be simplified by noting that the gauge field
$A^-$ on the antibranes may be taken to be trivial, so that $\ch({\cal S}_{
L_N^{1/2}}^-)=2^{k-1}$, while $A^+$ may be chosen to ensure that the
appropriate degree component of the Chern character $\ch({\cal S}_{
L_N^{1/2}}^+)$ is non-vanishing so as to produce a non-zero integral over the
transverse directions $\tilde\Sigma'$. Because of the assumed properties of the
function $f(x)$, this integral always converges, and thereby simply produces a
constant density factor in (\ref{WZred}). This leaves only the first integral
of (\ref{WZred}), which is the standard Chern-Simons action for a single BPS
D$(p-2k)$-brane wrapping a worldvolume $\tilde\Sigma$ and with $U(1)$
Chan-Paton gauge bundle $\cal L$ in Type II superstring theory (the density
factor then yields the appropriate tension). Thus the topological action
(\ref{WZfinal}) correctly reproduces the charge formula for the D-branes
obtained via tachyon condensation from the bound state of higher-dimensional
brane-antibrane pairs. In this context, since Clifford superconnections can be
thought of as quantizations of ordinary connections \cite{Quillen1,BGV}, a BPS
D-brane may be regarded as the ``classical limit'' of a non-BPS brane-antibrane
system.

This construction can be generalized by including the non-abelian transverse
scalar fields (\ref{calXdef}), which are sections of $N\Sigma\otimes({\rm
End}(E^+)\oplus{\rm End}(E^-))$, and by replacing the complex line bundle $\cal
L$ in (\ref{EcalLcalS}) by a bundle $\cal R$ of rank
\beq
M=\frac{N-\ch_0(I)-\ch_0({\cal I})}{2^{k-1}} \ ,
\label{Mrankdef}\eeq
where $\ch_0$ is the rank function and $N=N^+=N^-$ (This of course requires a
quantization condition on the ranks of the bundles in (\ref{EcalLcalS}) in
multiples of $2^{k-1}$). Under the stated properties of the spin$^c$ bundles
${\cal S}_{ L_N^{1/2}}^\pm$ above, it is possible to choose an appropriate
configuration of the scalar fields $\cal X$ such that the charge formula
(\ref{WZfinal}) coincides with the non-abelian Chern-Simons action for a system
of $M$ BPS D$(p-2k)$-branes with $U(M)$ Chan-Paton gauge bundle $\cal R$ and
including the T-duality invariant modifications from the adjoint sections of
$N\tilde\Sigma\otimes{\rm End}\,{\cal R}$. This generalizes the result
(\ref{WZred}) to multiple branes and realizes the $M$ D$(p-2k)$-branes as
generalized instanton-like configurations via tachyon condensation. In
particular, it is possible to realize the Myers dielectric effect \cite{myers}
whereby $M$ D$(p-2k)$-branes expand into a D$(p-2k+2r)$-brane (with $M$ units
of worldvolume instanton-like density) in terms of $N$ $p$-$\overline{p}$ pairs
opening up into $2^{k-1}$ $(p+2r)$-$\overline{(p+2r)}$ pairs. Some details of
this construction can be found in \cite{lozano}. Alternatively, the non-abelian
dielectric couplings can be induced by adding the terms $\gamma_i\,\phi^i$ to
the tachyon profile (\ref{ABStachyon})~\cite{AGS}.

\newsection{Ramond-Ramond Couplings on Unstable D-Branes}

In this section we will derive the Ramond-Ramond couplings on systems of
non-BPS D-branes in Type II superstring theory. We will present two
complimentary derivations of these actions. The first one is based on the old
geometric approach to Higgs fields through dimensional reduction \cite{dimred}.
In its simplest setting this technique introduces a single extra flat,
translationally invariant dimension. The tachyon field is then regarded as the
component of the gauge field along the extra direction. The main drawback of
this approach is that the superconnection gauge field should not depend on the
auxilliary coordinate, so that some of the physical information is lost through
the dimensional reduction that is encoded in the modes associated to the extra
dimension. This problem is cured by a second derivation of the Chern-Simons
actions through a particular reduction of the superconnection couplings of the
previous section. While this second approach is geometrically appealing because
it puts all of the non-supersymmetric configurations of D-branes in Type II
superstring theory into a common mathematical framework, it is the dimensional
reduction mechanism that plays the role in processes involving tachyon
condensation.

\subsection{Dimensional Reduction}

In this subsection we will derive the result for Type IIB D-branes and extend
it to the Type IIA case via T-duality. Consider a system of $N$ coincident
non-BPS D$p$-branes, with $p$ even. The mathematical description of this system
is much different than that of the brane-antibrane system, mainly because the
low-energy field content is drastically altered. As we will now demonstrate,
one way to think about this configuration is as the dimensional reduction of a
gauge theory in one higher dimension, rather than as a superconnection gauge
theory. This lends a somewhat different interpretation to the tachyon field
instability present in these systems.

The low-energy field content on the unstable system of branes consists of a
$U(N)$ gauge field $A_\mu$, and a Hermitian tachyon field $T$ which transforms
in the adjoint representation ${\bf N}\otimes\overline{\bf N}$ of the
Chan-Paton gauge group \cite{Sen4,Sen5,Sen3}. The field $A_\mu$ is a connection
of a $U(N)$ gauge bundle $E\to\Sigma$ over the $p+1$ dimensional worldvolume
$\Sigma$ of the $N$ unstable D$p$-branes. There is also a pair of massless,
16-component fermion fields $\psi_1,\psi_2$ which live in the adjoint
representation of $U(N)$ (coming from the massless Yang-Mills and tachyonic
supermultiplets). They are associated with the two possible Chan-Paton factors
carried by the open strings on each brane. The crucial issue concerns the
chiralities of these spinor fields under the local spacetime Lorentz group
$SO(9,1)$. In the static gauge, only an $SO(p,1)\times SO(9-p)$ subgroup of
$SO(9,1)$ is realized as a manifest symmetry of the worldvolume field theory.
Since $p$ is even, neither $SO(p,1)$ nor $SO(9-p)$ has a chiral spinor
representation, and the GSO projection cannot determine the $SO(9,1)$ chirality
of the fermion fields. Thus, both a left-handed and a right-handed
Majorana-Weyl spinor of $SO(9,1)$ will transform in the same spinor
representation of $SO(p,1)\times SO(9-p)$, even though the fermion zero modes
from the two Chan-Paton sectors of the open string spectrum on each D-brane
have the opposite GSO projection. Therefore, the low-energy worldvolume field
theory contains a pair of fermion fields $\psi_1,\psi_2$ each transforming in,
say, the right-handed Majorana-Weyl spinor representation of $SO(9,1)$.

Let us now consider the dimensional extension of the worldvolume $\Sigma$ to
the $p+2$ dimensional manifold
\beq
\hat\Sigma=\Sigma\times{\bf S}^1
\label{hatSigmadef}\eeq
and coordinatize the circle ${\bf S}^1$ by $y\in[0,1]$. The pair $(A_\mu,T)$
may then be thought of as the dimensional reduction to $\Sigma$ of a $U(N)$
gauge field $\hat A_M$ on $\hat\Sigma$ \cite{Horava},
\beq
\hat A_M=(A_\mu\,,\,T)~~~~~~,~~~~~~M=0,1,\dots,p,y \ .
\label{dimredAM}\eeq
The field (\ref{dimredAM}) may be regarded as a connection of a $U(N)$ gauge
bundle $\hat E_\rho\to\hat\Sigma$, where $\rho={\bf N}\otimes\overline{\bf N}$
is the $U(N)$ representation carried by the fermionic open string zero modes.
Thus the tachyon field $T$ on a system of non-BPS D-branes may be regarded as a
gauge connection of the external space ${\bf S}^1$, induced by a sort of
Kaluza-Klein mechanism from the reduction $\Sigma\times{\bf S}^1\to\Sigma$.
This is to be contrasted with the brane-antibrane system, in which the tachyon
field was regarded as a gauge connection of the discrete internal space
$\zed_2$, arising from a sort of Kaluza-Klein mechanism from the reduction
$\Sigma\times\zed_2\to\Sigma$.

The pair of spinor fields $\psi_1,\psi_2$ may be likewise regarded as the
dimensional reduction of a 32-component Majorana fermion field on $\hat\Sigma$,
\beq
\hat\Psi=\pmatrix{\psi_1\cr\psi_2\cr} \ .
\label{hatPsidef}\eeq
Therefore, the perturbative chiral gauge anomaly of the worldvolume field
theory on $\Sigma$ can be obtained from that of the oxidized theory on
$\hat\Sigma$ by dimensional reduction. In the latter theory, the only anomaly
that can arise is due to the massless chiral fermions, and applying the
standard index theorem as before we get
\beq
\ind\,i\hat\Dirac=(-1)^{(p+2)(p+3)/2}\int\limits_{\hat\Sigma}\ch^+\left
(\hat{\cal E}\right)\wedge\frac{{\rm Td}\left(T\hat\Sigma\otimes\complex
\right)}{\chi\left(T\hat\Sigma\right)} \ .
\label{indhatDirac}\eeq
Here the superbundle $\hat{\cal E}=\hat{\cal E}^+\oplus\hat{\cal E}^-$ is
defined by $\hat{\cal E}^\pm={\cal S}^\pm(\hat\Sigma)\otimes\hat E_\rho$, and
the Dirac operator $\hat\Dirac$ defines the two-term complex
\beq
 C^\infty\left(\hat\Sigma\,,\,\hat{\cal E}^\pm\right)~
\stackrel{\hat\Dirac}{\longrightarrow}~
 C^\infty\left(\hat\Sigma\,,\,\hat{\cal E}^\mp\right) \ .
\label{hat2term}\eeq
The characteristic classes in (\ref{indhatDirac}) may be simplified as
described in section 3.1 to yield
\beq
\ind\,i\hat\Dirac=\int\limits_{\hat\Sigma}\ch_\rho^+\left(\hat
E\right)\wedge\e^{d\bigl(N\hat\Sigma\bigr)}\wedge\frac{\widehat{A}\left(T\hat\Sigma
\right)}{\widehat{A}\left(N\hat\Sigma\right)}
\wedge\chi\left(N\hat\Sigma\right) \ .
\label{hatcharclasses}\eeq
The bundle $\hat E$ is trivially $\zed_2$-graded, and its Chern character may
be represented by a closed differential form on $\hat\Sigma$,
\beq
\ch^+_\rho\left(\hat E\right)=\ch_\rho\left(\hat E\right)=
\Tr_\rho\,\exp\frac1{2\pi i}\,\hat F_{\hat A} \ ,
\label{chEtriv}\eeq
where
\beq
\hat F_{\hat A}=d\hat A+\left[\hat A\,\stackrel{\wedge}{,}\,\hat A\right]
\label{hatFhatA}\eeq
is the field strength of the gauge field (\ref{dimredAM}).

The circle ${\bf S}^1$ is parallelizable in $X$, i.e. both its tangent and
normal bundles are trivial, so that $\chi(N{\bf S}^1)=\widehat{A}(T{\bf
S}^1)=\widehat{A}(N{\bf S}^1)=1$ and $d(N{\bf S}^1)=0$. Using the
multiplicativity (resp. additivity) of the characteristic classes
$\widehat{A}(V)$, $\chi(V)$ (resp. $d(V)$), and the decompositions
$T\hat\Sigma=T\Sigma\oplus T{\bf S}^1$ and $N\hat\Sigma=N\Sigma\oplus N{\bf
S}^1$, we find $\widehat{A}(T\hat\Sigma)=\widehat{A}(T\Sigma)$, and so on. It
is now straightforward to dimensionally reduce the index integral
(\ref{hatcharclasses}) to the D-brane worldvolume $\Sigma$. We expand the
fields on $\hat\Sigma$ in Fourier series around the ${\bf S}^1$,
\bea
\hat A_M(x,y)&=&\sum_{n=-\infty}^\infty A_M^{(n)}(x)~\e^{2\pi iny} \ , \nn\\
\hat\Psi(x,y)&=&\sum_{n=-\infty}^\infty\pmatrix{\psi_1^{(n)}(x)\cr
\psi_2^{(n)}(x)\cr}~\e^{2\pi iny} \ ,
\label{Fourier}\eea
which upon integrating over the ${\bf S}^1$ part of the integral
(\ref{hatcharclasses}) will localize the fields onto their zero modes
$A_\mu^{(0)}(x)=A_\mu(x)$, $A_y^{(0)}(x)=T(x)$, and
$\psi_a^{(0)}(x)=\psi_a(x)$, $a=1,2$, on $\Sigma$. In particular, the curvature
two-form (\ref{hatFhatA}) upon dimensionally reducing the fields becomes
\beq
\hat F_{\hat A}=F_A+D_AT\wedge dy \ ,
\label{hatFhatAred}\eeq
where $F_A$ is the field strength tensor of the original worldvolume gauge
field $A_\mu$ and
\beq
D_AT=dT+[A\,,\,T]
\label{DAT}\eeq
is the gauge covariant derivative of the tachyon field. Using (\ref{chEtriv})
and (\ref{hatFhatAred}), the ${\bf S}^1$ integration in (\ref{hatcharclasses})
may be carried out explicitly to give
\bea
\oint\limits_{{\bf S}^1}\ch_\rho^+\left(\hat
E\right)&=&\sum_{k=1}^\infty\frac1{(2\pi i)^k}\,\frac1{k!}\,\oint\limits_{{\bf
S}^1}\Tr_\rho\Bigl(F_A+D_AT\wedge
dy\Bigr)^k\nn\\&=&\sum_{k=1}^\infty\frac1{(2\pi i)^k}\,\frac1{(k-1)!}\,
\Tr_\rho\left((F_A)^{k-1}\wedge D_AT\right)\nn\\
&=&\frac1{2\pi i}~\Tr_\rho\left(D_AT\wedge\exp\frac1{2\pi i}\,F_A\right) \ .
\label{S1int}\eea
By using the Bianchi identity $D_AF_A=0$, we arrive finally at
\beq
\ind\,i\hat\Dirac=\frac1{2\pi i}\,\int\limits_\Sigma
d\,\Tr_\rho\left(T\,\exp\frac1{2\pi i}\,F_A\right)\wedge\e^{d(N\Sigma)}
\wedge\frac{\widehat{A}(T\Sigma)}
{\widehat{A}(N\Sigma)}\wedge\chi(N\Sigma) \ .
\label{indhatfinal}\eeq

As in section 3.1, we may readily argue that the quantum anomaly arising from
(\ref{indhatfinal}) can be cancelled by anomalous magnetic RR interactions.
Comparing with (\ref{calYfinal}) and incorporating the appropriate
modifications described in section 3.2, we arrive at the final form of the
Chern-Simons action describing the anomalous coupling of a system of $N$
non-BPS D-branes to the Ramond-Ramond fields $C$,
\bea
\widetilde{Z}_0&=&-\frac1{2\sqrt{2\pi}}\,\int\limits_\Sigma
{}~\Tr\,\phi^*\left(\e^{i(\jmath^{~}_\phi)^2-[\jmath^{~}_\phi,T]}~
\e^{\jmath^{~}_\phi d_\perp}
\biggm|_{x^\perp=0}\,\sum_{p=0}^{4+\sigma}C_{(2p+1-\sigma)}\wedge
\e^{-B/2\pi i}\right)\nn\\&&\wedge\,D_AT\wedge\exp\frac1{2\pi i}\,F_A\wedge
\sqrt{\widehat{A}(R_T)/\widehat{A}(R_N)}\wedge\e^{d(N\Sigma)/2} \ ,
\label{SWZnonBPS}\eea
where here Tr denotes the (symmetrized) trace in the fundamental representation
of the Chan-Paton gauge group $U(N)$. The $N\times N$ Hermitian matrices
$\phi^i$ describe the transverse degrees of freedom of the non-BPS branes, and
the commutator term $[\jmath^{~}_\phi,T]$ arises from the dimensional reduction
of the pull-back modification analogous to (\ref{pullbacks},\ref{nablaNdef})
involving the gauge covariant derivative $\hat D_{\hat A}$. The subscript 0 on
the action (\ref{SWZnonBPS}) emphasizes the fact that it contains only the zero
modes of the tachyon and gauge fields from the dimensional reduction, since in
the calculation above we have simply eliminated the ${\bf S}^1$ dependence of
all fields before integrating over the extra dimension. A more precise
evaluation should keep all higher Kaluza-Klein modes in (\ref{Fourier}) before
integrating over the circle. However, in the present approach, it is difficult
to keep track of these higher excitations, given that the index theory
calculation relies on the structure of the lowest lying modes. The higher
fermion modes, for example, have masses of order $n^2$ for the $n$-th
Kaluza-Klein state, and the appropriate anomaly cannot be identified for these
massive fields, in the present energy regime that the calculations are based
on. Furthermore, there is no immediate interpretation of these higher states in
the original worldvolume field theory. To correctly account for the rest of the
Kaluza-Klein spectrum, we will present another calculation of the anomalous
coupling of non-BPS D-branes, which utilizes the previous superconnection
formalism. This lends a more precise interpretation of the tachyon field which
is based on the previous constructions.

\subsection{Reduction from Brane-Antibrane Pairs}

In this subsection we will begin by working in Type IIA superstring theory. An
unstable IIA(B) D$p$-brane may be realized as the projection of a IIB(A)
D$p$-D$\overline{p}$ system by the discrete $\zed_2$ symmetry generated by the
operator $\klein$ \cite{Sen4,BHYi}, where $F_{\rm L}$ is the left-moving part
of the spacetime fermion number operator. The operator $\klein$ acts as
multiplication by $-1$ on all Ramond sector states in the left-moving part of
the fundamental string worldsheet, leaving all other sectors unchanged. It
exchanges a D-brane with its antibrane, so that a brane-antibrane pair is
invariant under $\klein$ and it makes sense to take the $\zed_2$ quotient of
this configuration. The feature that the $\klein$ projection maps Type IIB
superstring theory into Type IIA superstring theory can be proven using
boundary states \cite{DasPark}. The fact that the brane-antibrane pair is
mapped to a non-BPS D-brane follows from the action of the operator $\klein$ on
the Chan-Paton factors $\psi^{~}_{\rm CP}\in U(2N)$ of the open strings in the
D$p$-D$\overline{p}$ system \cite{Sen4}, which is given by
\beq
\klein\,:\,\psi^{~}_{\rm CP}~\longmapsto~\sigma_1\,\psi^{~}_{\rm CP}
\,\sigma_1 \ ,
\label{KleinCP}\eeq
where
\beq
\sigma_1=\pmatrix{0&\id_N\cr\id_N&0\cr}
\label{sigmaCliff}\eeq
generates the one-dimensional complex Clifford algebra
$\cliff_1^*=\complex\oplus\complex\sigma_1$.

As we discussed earlier, the lowest lying (GSO projected) bosonic states of a
system of $N$ brane-antibrane pairs wrapping a common worldvolume $\Sigma$ may
be geometrically encoded in the superconnection
\beq
\hat\SA=d+\hat{\cal A}=\pmatrix{d+\hat A^+&\hat T^\dagger\cr\hat T&d+
\hat A^-\cr}
\label{hatsuperconn}\eeq
where $\hat A^\pm$ are the $U(N)$ gauge fields on the branes and antibranes,
respectively, and $\hat T$ is the bi-fundamental tachyon field from the open
string $p$-$\overline{p}$ states. The curvature of the superconnection
(\ref{hatsuperconn}) is
\beq
\hat F_{\hat\sSA}=\pmatrix{\hat F^++\hat T^\dagger\hat T&
\hat D\hat T^\dagger\cr\hat D\hat T&\hat F^-+\hat T\hat T^\dagger\cr} \ .
\label{hatsuperconncurv}\eeq
Upon introducing the fields
\bea
A&=&\frac12\,\left(\hat A^++\hat A^-\right) \ , \nn\\
\bar A&=&\frac12\,\left(\hat A^+-\hat A^-\right) \ ,\nn\\
T&=&\frac12\left(\hat T+\hat T^\dagger\right) \ , \nn\\
\bar T&=&\frac12\left(\hat T-\hat T^\dagger\right) \ ,
\label{ppbarredfields}\eea
it follows that the only ones which survive the $\klein$ projection are $A$ and
$T$, i.e. the quotient of the spectrum of the $p$-$\overline{p}$ pairs sets
$\bar A=\bar T=0$. Clearly, the fermionic spectrum of the quotiented theory
contains two Majorana-Weyl spinors of the same chirality, as the projection
identifies the field contents on the branes and antibranes. In this way we
recover the low-energy spectrum of fields on the worldvolume $\Sigma$ of a
system of $N$ non-BPS D$p$-branes.

We may therefore compute the anomalous coupling of the unstable D-branes by
taking the quotient of the anomaly term (\ref{indexfinal}) for the
brane-antibrane system. Evidently, the only change is the reduction of the
Chern character, which with $\bar A=\bar T=0$ in (\ref{hatsuperconncurv})
becomes
\beq
\left[\ch^+\left(\hat E\right)\right]_\klein
=\Tr~(-1)^F\,\exp\frac1{2\pi i}\,\pmatrix{F_A+T^2&D_AT\cr D_AT&F_A+T^2\cr} \ ,
\label{kleinchern}\eeq
where $\hat E=\hat E^+\oplus\hat E^-$ is the Chan-Paton superbundle over the
brane-antibrane worldvolume $\Sigma$, which reduces to a trivially
$\zed_2$-graded $U(N)$ gauge bundle $E\to\Sigma$ with corresponding worldvolume
fields in (\ref{kleinchern}) upon modding out by the operator $\klein$ using
its action (\ref{KleinCP}) on the Chan-Paton factors. This $\zed_2$-action
induces an isomorphism $\hat E^+\cong\hat E^-\equiv E$ and thereby identifies
$E$ as the diagonal sub-bundle of $[\hat E]_\klein\cong E\oplus E$. Using the
supersymmetric structure of (\ref{kleinchern}) and (\ref{gradingop}) with
$N^+=N^-=N$, we may diagonalize the real, symmetric reduced superconnection
field strength $[\hat F_{\hat\sSA}]_\klein$ to obtain
\beq
\left[\ch^+\left(\hat E\right)\right]_\klein
=\Tr\left[\exp\frac1{2\pi i}\,\Bigl(F_A+T^2+D_AT\Bigr)-\exp\frac1{2\pi i}\,
\Bigl(F_A+T^2-D_AT\Bigr)\right] \ .
\label{kleinchernred}\eeq
Note that an elegant reduction such as (\ref{kleinchernred}) is not possible
for the Chern character of a brane-antibrane system itself, whereby the
superconnection curvature is generically a complex, Hermitian matrix with
respect to the $\zed_2$-grading.

The reduced Chern character (\ref{kleinchernred}) can be expanded into a more
explicit expression by using the Dynkin form of the Baker-Campbell-Hausdorff
formula \cite{vara}. This enables us to write
\beq
\exp\frac1{2\pi i}\,\Bigl(F_A+T^2+D_AT\Bigr)=\exp\left[\frac1{2\pi i}\,
\Bigl(T^2+D_AT\Bigr)+\sum_{r,s=1}^\infty{\Xi}_{rs}[A,T]\right]\wedge
\exp\frac1{2\pi i}\,F_A
\label{DynkinBCH}\eeq
where
\beq\new{\begin{array}{c}
{\Xi}_{rs}[A,T]=\frac{(-1)^s}{(r+s)(2\pi
i)^{r+s}}\,\sum_{m\geq1}\frac{(-1)^m}m\,\sum_{\stackrel{p_1+q_1\geq1,\dots,p_m+q_m\geq1}
{p_1+\dots+p_m=r\,,\,q_1+\dots+q_m=s}}\,\frac1{p_1!q_1!\cdots p_m!q_m!}\nn\\
\times\,\Bigl[(F_A+T^2+D_AT)^{p_1}\,\stackrel{\wedge}{,}\,\Bigl[(F_A)^{q_1}\,
\stackrel{\wedge}{,}\,\Bigl[\cdots\Bigl[(F_A+T^2+D_AT)^{p_m}\,
\stackrel{\wedge}{,}\,(F_A)^{q_m}\Bigr]~\Bigr]~\Bigr]\cdots\Bigr] \ .
\end{array}}
\label{calCrsdef}\eeq
The leading terms in the expansion of the right-hand side of (\ref{DynkinBCH})
are given by
\bea
{\Xi}_2[A,T]&=&-\frac1{2(2\pi i)^2}\,\Bigl[F_A\,\stackrel{\wedge}{,}\,
F_A+T^2+D_AT\Bigr] \ , \nn\\{\Xi}_3[A,T]&=&\frac1{12(2\pi i)^3}\,
\left(\Bigl[~\Bigl[F_A+T^2+D_AT\,\stackrel{\wedge}{,}\,F_A\Bigr]\,
\stackrel{\wedge}{,}\,F_A\Bigr]\right.\nn\\& &+\left.\Bigl[~\Bigl[F_A+
T^2+D_AT\,\stackrel{\wedge}{,}\,F_A\Bigr]\,\stackrel{\wedge}{,}\,
F_A+T^2+D_AT\Bigr]\right) \ , \nn\\{\Xi}_4[A,T]&=&-\frac1{48(2\pi i)^4}\,
\left(\Bigl[F_A\,\stackrel{\wedge}{,}\,\Bigl[F_A+T^2+D_AT\,
\stackrel{\wedge}{,}\,\Bigl[F_A+T^2+D_AT\,\stackrel{\wedge}{,}\,F_A
\Bigr]~\Bigr]~\Bigr]\right.\nn\\& &+\left.\Bigl[F_A+T^2+D_AT\,
\stackrel{\wedge}{,}\,\Bigl[F_A\,\stackrel{\wedge}{,}\,\Bigl[F_A+T^2+D_AT\,
\stackrel{\wedge}{,}\,F_A\Bigr]~\Bigr]~\Bigr]\right) \ , \\&\vdots&\nn
\label{calClow}\eea
where ${\Xi}_n[A,T]=\sum_{r+s=n}{\Xi}_{rs}[A,T]$. Substituting
(\ref{DynkinBCH}) into (\ref{kleinchernred}), we arrive, in the usual way, at
the complete RR coupling
\bea
\widetilde{Z}&=&-\frac{\sqrt{2\pi}}2\,\int\limits_\Sigma
{}~\Tr\,\phi^*\left(\e^{i(\jmath^{~}_\phi)^2-[\jmath^{~}_\phi,T]}
{}~\e^{\jmath^{~}_\phi d_\perp}
\biggm|_{x^\perp=0}\,\sum_{p=0}^{4+\sigma}C_{(2p+1-\sigma)}\wedge\e^{-B/2\pi i}
\right)\nn\\& &\wedge\left\{\exp\left(\frac1{2\pi i}\,
\Bigl(T^2+D_AT\Bigr)+\sum_{r,s=1}^\infty{\Xi}_{rs}[A,T]\right)
\right.\nn\\& &-\left.\exp\left(\frac1{2\pi i}\,\Bigl(T^2-D_AT
\Bigr)+\sum_{r,s=1}^\infty{\Xi}_{rs}[A,-T]\right)\right\}\wedge
\exp\frac1{2\pi i}\,F_A\nn\\& &\wedge\,\sqrt{\widehat{A}(R_T)/
\widehat{A}(R_N)}\wedge\e^{d(N\Sigma)/2} \ ,
\label{tildeS0compl}\eea
where $\phi^i=\frac12\,(\hat\phi_+^i+\hat\phi_-^i)$ and now the transverse
tachyon coupling $[\jmath^{~}_\phi,T]$ comes from the $\klein$ projection of
the operator (\ref{calJcalXT}). The action (\ref{tildeS0compl}) is an odd
function of the tachyon field $T$. Specifically, it contains only even powers
of $T$ and odd powers of $D_AT$. The expansion of (\ref{tildeS0compl}) is
similar in form to that constructed in \cite{kluson1}, except that it
generically contains extra powers of the field strength $F_A$ coupled to the
tachyon terms. To linear order in the tachyon field, the action
(\ref{tildeS0compl}) coincides with the zero mode action (\ref{SWZnonBPS}) and
hence the non-BPS D-brane coupling proposed in \cite{bcr}. The complete series
(\ref{tildeS0compl}) thereby represents the contributions from all Kaluza-Klein
sectors of the oxidized theory described in the previous subsection.

\subsection{Tachyon Condensation}

Using the action (\ref{Skindef}) it is possible to write down a natural
geometric action for the system of unstable D-branes using the $\klein$
projection. It is given by
\bea
\widetilde{Z}_{\rm kin}&=&\frac12\,\int\limits_\Sigma\left[
\left\|\hat F_{\hat\sSA}-\hat{\cal G}\right\|^2\right]_\klein
\nonumber\\&=&\int\limits_\Sigma
d{\rm vol}(\Sigma)~\Tr\left[\left(F_A\right)^2+\left(D_AT\right)^2+
\left(T^2-{\cal G}\right)^2+\dots\right] \ ,
\label{Zkinunstable}\eea
where ${\cal G}=\frac12\,(\hat{\cal G}^++\hat{\cal G}^-)$. The resulting
tachyon potential in (\ref{Zkinunstable}) has the anticipated $\zed_2$
reflection symmetry under the transformation $T\mapsto-T$ \cite{osrev}. Again
the action (\ref{Zkinunstable}) is minimized by flat gauge connections and
covariantly constant tachyon fields. If ${\cal G}=m^2\,\id$ with $m^2>0$, then
the tachyon condensates obey the equation $T_c^2=m^2\,\id$ and the $U(N)$ gauge
symmetry of the system is broken down to the subgroup $U(n_c)\times U(N-n_c)$,
where $n_c$ is the number of negative eigenvalues of $T_c$. Again this action
is determined by a spectral Lagrangian of the form $\Tr[\sigma\circ
c^{-1}(\hat\Sdirac^2)]^2_\klein$ involving the pertinent generalized Dirac
operator. Thus for non-BPS D-branes, the same geometrical ingredients naturally
lead to the standard processes involving tachyon condensation.

In the absence of a tachyon field the RR couplings (\ref{tildeS0compl}) vanish,
as expected since then the unstable D-brane configuration simply decays into
the supersymmetric vacuum state. However, a topologically non-trivial tachyonic
configuration can produce a lower dimensional D-brane within the configuration
of non-BPS branes. To demonstrate this, the crucial observation is that for
processes involving tachyon condensation it is sufficient to focus on the zero
mode part (\ref{SWZnonBPS}) of the total Chern-Simons action \cite{kluson1}.
For a Higgs profile of the tachyon field, the terms in (\ref{tildeS0compl})
involving higher powers of $T$ or $D_AT$ will vanish. By dropping the
non-abelian couplings to the transverse scalar fields, the action
(\ref{tildeS0compl}) coincides with that of \cite{bcr}. It is now
straightforward to repeat the brane construction of section~3.3 in the present
case and induce the Chern-Simons term for a BPS D-brane wrapping a worldvolume
$\tilde\Sigma$ of codimension $2k+1$ in $\Sigma$. For a Higgs-like
configuration, the action (\ref{tildeS0compl}) can be reduced to the form
\bea
\widetilde{Z}_0&=&-\frac1{2\sqrt{2\pi}}\,\int\limits_{\tilde\Sigma}~
\sum_{p=0}^{4+\sigma}\phi^*\left(C_{(2p+1-\sigma)}\wedge\e^{-B/2\pi i}\right)
\wedge\ch({\cal L})\wedge\sqrt{\frac{\widehat{A}\left(T\tilde\Sigma\right)}
{\widehat{A}\left(N\tilde\Sigma\right)}}\wedge\e^{d\bigl(N\tilde\Sigma
\bigr)/2}\nonumber\\&&\times\,\int\limits_{\tilde\Sigma'}d\,\Tr\left(T\exp
\frac1{2\pi i}\,F_A\right)\wedge\widehat{A}\left[N\left(\tilde\Sigma\,,\,
\Sigma\right)\right]\wedge\e^{-c_1( L_N)/2} \ ,
\label{WZredunstable}\eea
where we have used the Bianchi identity. Here $A$ is a connection on the
twisted real spinor bundle ${\cal S}_{ L_N^{1/2}}$ over $\tilde\Sigma$, while
for the tachyon field configuration we have again taken Clifford multiplication
\beq
T(x)=\id_{\cal L}\otimes\left(f(x)\,\sum_{i=1}^{2k+1}\Gamma_i\,x^i\right)
\oplus\id_I \ ,
\label{ABStachyonunstable}\eeq
with $\Gamma_i$ the generators of the Clifford algebra of the transverse
structure group $SO(2k+1)$. From (\ref{WZredunstable}) it is evident that an
appropriate choice of gauge connection $A$ leads to the correct Ramond-Ramond
coupling of a supersymmetric D$(p-2k-1)$-brane, similarly to \cite{bcr}. Since
the characteristic classes are closed forms, the second integral in
(\ref{WZredunstable}) can be reduced to the form
$\oint_{\partial\tilde\Sigma'}\Tr(T\,\e^{F_A/2\pi
i})\wedge\widehat{A}\wedge\e^{-c_1/2}$. Since the tachyon field $T$ is constant
on $\partial\tilde\Sigma'$, we can choose $A$ so that its field strength $F_A$
has the appropriate generalized vortex configuration to yield a non-vanishing
boundary integration.

\newsection{Other Non-Supersymmetric Brane Systems}

In this section we will briefly explain how to obtain the anomalous couplings
to all non-BPS systems of branes in Type II superstring theory. We shall do so
by giving a set of rules for the transformations of the RR potentials in our
previously derived Chern-Simons actions.

\subsection{NS-Branes}

An important ingredient missing from our analysis of the Type IIB theory is its
S-duality symmetry. This duality is also manifest at the level of $p$-brane
solutions of ten-dimensional supergravity. The Chern-Simons actions for
unstable configurations of NS-branes in Type IIB superstring theory can be read
off from the couplings (\ref{WZfinal}) and (\ref{tildeS0compl}) by applying the
$SL(2,\zed)$ S-duality transformation rules \cite{BEHvdSHL} to the total RR
potential ${\cal C}=\sum_p\phi^*C_{(p)}$, the $B$-field, and the field
strengths $F^\pm$ of the open fundamental strings ending on the D-branes. In
the string frame, these transformations are given by
\bea
C_{(0)}&\longmapsto&-\frac{C_{(0)}}{\left(C_{(0)}\right)^2+\e^{-2\varphi}} \ ,
\nn\\C_{(2)}&\longmapsto&B \ , \nn\\C_{(4)}&\longmapsto&C_{(4)} \ , \nn\\
C_{(6)}&\longmapsto&B_{(6)} \ , \nn\\C_{(8)}&\longmapsto&-\widetilde{C}_{(8)}
\ , \nn\\B&\longmapsto&-C_{(2)} \ , \nn\\F^\pm&\longmapsto&
\widetilde{F}^\pm \ .
\label{DNSrules}\eea
Here $\varphi$ is the dilaton field, $C_{(0)}$ the axion field, $B_{(6)}$ is
the electromagnetic dual of the NS-NS two-form $B$, and $\widetilde{F}^\pm$ are
the fluxes of the D1-branes that end on the NS-branes, which combine
geometrically with the induced open $p$-$\overline{p}$ D-string tachyon field
$\widetilde{T}$ into the appropriate superconnection on the NS-brane
worldvolume $\widetilde{\Sigma}$. Note that the four-form RR potential is
unaffected because of the self-duality of the D3-brane, while the
transformation of the eight-form potential reflects the fact that D7-branes and
NS7-branes do not form a doublet under S-duality \cite{EL}. The eight-form
$\widetilde{C}_{(8)}$ is related to the NS-NS and RR eight-forms by
\beq
d\widetilde{C}_{(8)}=-C_{(0)}~dB_{(8)}+\left[\left(C_{(0)}\right)^2+
\e^{-2\varphi}\right]~dC_{(8)} \ .
\label{tildeC8B8C8}\eeq
The fields $(C_{(8)},\widetilde{C}_{(8)},B_{(8)})$ thereby form a triplet under
$SL(2,\zed)$ transformations, where in addition to the transformation rule in
(\ref{DNSrules}) we have
\bea
\widetilde{C}_{(8)}&\longmapsto&-C_{(8)} \ , \nn\\
B_{(8)}&\longmapsto&-B_{(8)} \ .
\label{tripletmap}\eea
Via T-duality, we also recover in this way the corresponding Chern-Simons
actions for Type IIA NS-branes. The bound state constructions of BPS NS-branes
from unstable ones now also follow as outlined in sections 3.3 and 4.3. Some
details can be found in~\cite{lozano,HL1,HL2}.

\subsection{M-Branes}

The constructions of previous sections yield the anomalous couplings of all
branes in Type II superstring theory, with the exception of the gravitational
wave and the Kaluza-Klein monopole which are only defined in spacetimes that
contain a special isometric direction. In the case of the pp-wave this isometry
lies in the direction of propagation of the wave, while for the KK-monopole it
corresponds to the Taub-NUT fiber of its normal bundle. By oxidizing Type IIA
superstring theory to M-Theory, these solitonic branes can be most naturally
seen to arise from reductions of the corresponding M-branes in 11 dimensions.
Nine-branes and ten-branes in M-Theory should, however, be dealt with in the
context of {\it massive} 11-dimensional supergravity \cite{BLO}, since the BPS
M9-brane couples magnetically to the mass field. While a fully covariant
massive supergravity theory cannot be constructed in 11-dimensions \cite{BDHS},
a supergravity action that is gauged with respect to an isometric 11-th
direction of spacetime can be written down which reduces dimensionally to
massive Type IIA Romans supergravity \cite{Romans}. Reduction of the single
M9-brane in this way along its gauged direction yields a D8-brane domain wall,
while reduction along the transverse direction gives an NS9-brane. Reduction in
another direction produces a KK8-brane monopole \cite{EL} with a gauged
direction in its worldvolume that is inherited from the M9-brane. In turn,
these latter branes are most naturally understood as the electromagnetic duals
of the mass field in massive Type IIA supergravity. Similarly, reduction of an
M0-brane along the direction of its Killing vector yields a D0-brane, while its
reduction in a non-isometric direction produces a gravitational wave.

The rules for uplifting the unstable Type IIA Chern-Simons actions
(\ref{WZfinal}) and (\ref{tildeS0compl}) along the direction of a Killing
vector field $\hat{k}$ to M-Theory actions describing the couplings of unstable
M-branes are given by \cite{lozano,BEL}
\bea
C_{(1)}&\longmapsto&\frac{\hat{k}_{(1)}}{\left\|\hat{k}\right\|^2} \ ,
\nn\\C_{(3)}&\longmapsto&\left(\widehat{C}_{(3)}\right)_{\hat\mu\hat\nu
\hat\lambda}~\nabla^{\hat k}\widehat{\cal X}^{\,\hat\mu}\wedge
\nabla^{\hat k}\widehat{\cal X}^{\,\hat\nu}\wedge
\nabla^{\hat k}\widehat{\cal X}^{\,\hat\lambda} \ , \nn\\
C_{(5)}&\longmapsto&\jmath^{~}_{\hat k}\widehat{C}_{(6)}+\widehat{C}_{(3)}
\wedge\jmath^{~}_{\hat k}\widehat{C}_{(3)} \ , \nn\\C_{(7)}&\longmapsto&
\jmath^{~}_{\hat k}\widehat{N}_{(8)} \ , \nn\\C_{(9)}&\longmapsto&
\jmath^{~}_{\hat k}\widehat{B}_{(10)} \ , \nn\\B&\longmapsto&
\jmath^{~}_{\hat k}\widehat{C}_{(3)} \ ,
\nn\\F^\pm&\longmapsto&\widehat{F}^\pm \ .
\label{Muplifts}\eea
The $p$-form fields on the right-hand side of (\ref{Muplifts}) are assumed to
be invariant under the isometry of spacetime, ${\cal L}_{\hat k}\widehat{C}=0$
where ${\cal L}_{\hat k}$ is the Lie derivative along the Killing vector field
$\hat k$, and the hats refer to 11-dimensional quantities. The one-form $\hat
k_{(1)}$ is the Poincar\'e-Hodge dual to the Killing vector field, while
$\widehat{C}_{(3)}$ (resp. $\widehat{C}_{(6)}$) is the usual three-form (resp.
six-form) field of 11-dimensional supergravity. The matrix fields
$\widehat{\cal X}^{\,\hat\mu}$ are the non-abelian M-brane embedding
coordinates in 11-dimensions, while $\nabla^{\hat k}$ is the usual non-abelian
M-brane covariant derivative defined by
\beq
\nabla^{\hat k}\widehat{\cal X}^{\,\hat\mu}=\widehat{D}
\widehat{\cal X}^{\,\hat\mu}-\frac{\hat k_{\hat\nu}\,\widehat{D}
\widehat{\cal X}^{\,\hat\nu}}{\left\|\hat{k}\right\|^2}\,\hat k^{\hat\mu}
\label{nablahatk}\eeq
where $\widehat{D}$ is the 11-dimensional gauge covariant derivative defined
analogously to (\ref{DTdef}). The eight-form field $\widehat{N}_{(8)}$ is the
Hodge dual of the Killing one-form $\hat k_{(1)}$, while $\widehat{B}_{(10)}$
is the electromagnetic dual of the mass field. The gauge field curvatures
$\widehat{F}^\pm$ are the fluxes of the M2-branes wrapped around the direction
of $\hat k$, which induce a $p$-$\overline{p}$ tachyon field $\widehat{T}$ in
the M-brane worldvolume $\widehat{\Sigma}$ that combine into the appropriate
superconnection. The KK-monopole and pp-wave couplings can now be obtained by
oxidizing the appropriate Type IIA actions using the transformation rules
(\ref{Muplifts}), and then dimensionally reducing them in a worldvolume
direction. The corresponding transformations of all objects appearing in the
Chern-Simons actions (\ref{WZfinal}) and (\ref{tildeS0compl}) can be worked out
in the same way as the other dimensional reductions discussed in this paper.
Further details can be found in \cite{lozano,HL1,HL3}, where the corresponding
bound state constructions are also given. In fact, all Type II branes can be
obtained via the totality of reductions and tachyon condensations on the
oxidized M-brane couplings described in this subsection \cite{lozano,HL1}.

\newsection{K-Theory Analysis}

Let $\Sigma$ be a compact spin$^c$ brane-antibrane worldvolume manifold in Type
II superstring theory, and let $E=E^+\oplus E^-$ be the corresponding
Chan-Paton superbundle over $\Sigma$. Under the usual physical assumptions of
brane-antibrane creation and annihilation
\cite{NonBPSrev,Sen1,WittenK,osrev,Sen2}, the details of which have been
substantially verified via boundary string field theory calculations
\cite{BSFT1}, the Chan-Paton bundles on the branes and antibranes should be
subjected to the equivalence relation $(E^+,E^-)\sim(E^+\oplus H,E^-\oplus H)$
for all gauge bundles $H$, and hence the net D-brane charge of this
configuration depends only on its K-theory class
$[(E^+,E^-)]=[E^+]-[E^-]\in\K^0(\Sigma)$. By using the Thom isomorphism and the
Atiyah-Bott-Shapiro construction~\cite{WittenK,osrev}, it is possible to map
this class to an element of the K-theory group $\K^0(X)$ of spacetime. This
utilizes the construction that was presented in section~3.3 which derived
D-brane charge from a higher-dimensional brane-antibrane system. In this
subsection we will show how this fact ties in very naturally with the
superconnection formalism developed in this paper in terms of Dirac operators
and index theory. This identification is supported by the analysis of the
previous section, which shows that {\it all} branes are universally classified
by the appropriate K-theory groups~\cite{HL2,HL3}.

An interpretation of the anomalous coupling in
(\ref{Thetaanomdef},\ref{WZansatz}) for supersymmetric systems of D-branes,
which also takes into account the subtleties associated with the self-duality
of the RR fields~\cite{MooreWitten}, has been given in terms of K-theory
in~\cite{FreedHop}. One can define a K-theory group $\hat\K^{p+1}(X)$ of
$p$-form fields, analogously to the framework of Deligne cohomology, via the
exact sequence
\beq
0~\longrightarrow~\K^p(X,\real)\,/\,\K^p(X)~\longrightarrow~\hat\K^{p+1}(X)
{}~\longrightarrow~{\rm B}^{p+1}(X)~\longrightarrow~0 \ ,
\label{hatKexactseq}\eeq
where $\K^p(X,\real)=\K^p(X)\otimes\real$ and
\beq
{\rm B}^{p+1}(X)=\Bigl\{(x,\omega)\in\K^{p+1}(X)\times H^{p+1}(X,\zed)~\Bigm|
{}~\ch(x)=[\omega]^{~}_{\rm DR}\Bigr\} \ ,
\label{BXdef}\eeq
with $[\omega]^{~}_{\rm DR}\in H^{p+1}(X,\real)$ the de~Rham representative of
the integer cohomology class $\omega$. For a system of BPS D$p$-branes we then
have $\e^{iZ}\in\hat\K^{p+1}(X)$, where the type of K-theory (complex, real or
quaternionic) depends on the value of $p$ mod 8~\cite{FreedHop}. However, in
what follows we shall see that the analysis of the present paper is more
naturally connected to K-homology, showing that K-homology is really the
appropriate setting for the topological classification of D-brane charge. This
has been pointed out previously in different contexts, and from very different
points of view, in~\cite{osrev,KDhomology,NCtachyon}. We shall only deal with
the construction of K-theory classes over the worldvolume $\Sigma$, as then the
mapping to the K-theory of spacetime can be carried through by using standard
techniques.

\subsection{Index Bundles and Chern-Simons Couplings}

In this subsection we shall work in Type IIB superstring theory, so that
$\dim\Sigma=p+1$ is even. Consider the Clifford bundle $\cliff(\Sigma)$ over
$\Sigma$, and let $\dirac$ be the corresponding Dirac operator. By using Swan's
theorem, we can represent the Chan-Paton superbundle over $\Sigma$ as the range
$E=\Pi\,{\cal O}_N(\Sigma)$ of a projection $\Pi:{\cal O}_N(\Sigma)\to{\cal
O}_N(\Sigma)$, $\Pi^2=\Pi=\Pi^\dagger$, of the trivial vector bundle ${\cal
O}_N(\Sigma)\to\Sigma$ of rank $N=(p+1)\,\ch_0(E)$. The corresponding twisted
Dirac operator $\dirac_E$ on $\cliff(\Sigma)$ with coefficients in $E$ may
thereby be expressed as
\beq
\dirac_E=\Pi\,(\dirac\otimes\id)\,\Pi \ .
\label{diracPi}\eeq
With $\hil= C^\infty(\Sigma,S_E)=\hil^+\oplus\hil^-$ the graded Hilbert space
of smooth $E$-valued spinor fields on $\Sigma$, the Dirac operator is a map
$\dirac_E:\hil^\pm\to\hil^\mp$ and so it can be decomposed with respect to the
$\zed_2$-grading as
\beq
\dirac_E=\pmatrix{0&\dirac_E^-\cr\dirac_E^+&0\cr} \ .
\label{diracEdecomp}\eeq
If $\nabla$ is an ordinary connection on $E$, then the quantity
\beq
\SE=(-1)^F\otimes\nabla+\dirac_E=\pmatrix{\nabla&\dirac_E^-\cr\dirac_E^+&
-\nabla\cr}
\label{superconndirac}\eeq
defines a superconnection of the twisted spinor bundle $S_E=S_E^+\oplus S_E^-$.
The Chern character of $S_E$ is then given by~\cite{Quillen1}
\beq
\ch^+(S_E)=\ch^+(\hil,\dirac_E)=\Tr~\exp\frac1{2\pi i}\,F_\sSE
\label{SEChern}\eeq
where in the first equality we have emphasized the fact that $\ch^+$ depends
only on the choice of Dirac operator $\dirac_E$ acting on a particular graded
Hilbert space $\hil$. The formula (\ref{SEChern}) follows from the one-to-one
correspondence between generalized Dirac operators and superconnections that we
mentioned in section~2.3, and the Lichnerowicz formula (\ref{Lich}). Moreover,
the cohomology class of (\ref{SEChern}) does not depend on the off-diagonal odd
parts of the superconnection (\ref{superconndirac}), and so it determines an
element $\ch(E^+)-\ch(E^-)\in H^{\rm even}(\Sigma,{\bf Q})$. In other words,
the Chern character depends only on the choice of virtual bundle
$[(E^+,E^-)]\in\K^0(\Sigma)$. Going back to the superconnection $\SA$
introduced in section~2.1, this simply means that the induced D-brane charge is
independent of the choice of profile for the tachyon field $T$. This is
precisely what was found in section~3.3 via explicit calculation.

The key feature here is that the anomalous coupling on the brane-antibrane
worldvolume is determined entirely by the choice of Dirac operator, or more
precisely by the pair $({\cal H},\dirac_E)$. As we will now discuss, this
immediately leads to the relationship to K-theory, or more precisely to
K-homology. The family of finite-dimensional subspaces $\ker
i\dirac_E^\pm\subset\hil^\mp$ defines a virtual bundle over $\Sigma$ known as
the index bundle~\cite{osrev,Atiyahbook,Connesbook}
\beq
\Ind(\hil,\dirac_E)=\left[\ker i\dirac_E^+\right]-\left[\ker i\dirac_E^-\right]
\in\K^0(\Sigma) \ .
\label{indexbundle}\eeq
The closed differential form ({\ref{SEChern}) is then also a representative of
the Chern character $\ch^+(\Ind(\hil,\dirac_E))\in H^{\rm even}(\Sigma,{\bf
Q})$ of the index bundle. By using this property and the generic, untwisted
Dirac operator $\dirac$, we may define a natural pairing on K-theory known as
the index map
\beq
\Index_\dirac\,:\,\K^0(\Sigma)~\longrightarrow~\zed \ ,
\label{Indexmap}\eeq
which is given by
\bea
\Index_\dirac\Bigl([E]\Bigr)&=&\ind\,i\dirac_E\nn\\&=&\ch_0\Bigl(
\Ind(\hil,\dirac_E)\Bigr)\nn\\&=&\dim\ker i\dirac_E^+-\dim\ker i\dirac_E^- \ .
\label{Indexmapexpl}\eea
Therefore, the anomaly arising from the Dirac spinor fields on $\Sigma$ leads
naturally to a pairing on K-theory. This pairing in turn defines K-homology, as
we will discuss in the next subsection.

Before doing this, let us first show precisely how the index bundle is related
to the generalized Chern-Simons forms which are used to generate the
Ramond-Ramond couplings, and hence the appropriate pairing on K-theory. We
write the generalized Dirac operator $\dirac_E=Q(\nabla^{\rm
s}\otimes\id+\id\otimes\alg)\equiv\SlashN^{\rm s}+\SlashA$ as in section~2.3,
and introduce a linear homotopy of superconnections. For fixed $t\in[0,1]$, we
consider the superconnection
\beq
\SE(t)=\delta_{\rm BRST}+\SlashN^{\rm s}+t(\Lambda+\SlashA) \ ,
\label{curvhomotopy}\eeq
where $\Lambda$ is the Cartan-Maurer form (\ref{Cartanform}). Its curvature is
\beq
\SE(t)^2=(t^2-t)\Bigl(\Lambda^2+[\Lambda,\SlashA]\Bigr)+(\SlashN^{\rm s}
+t\SlashA)^2 \ ,
\label{SEtcurv}\eeq
which, at $t=0,1$, obeys the horizontality condition \cite{Quillen1}
\bea
\SE(0)^2&=&Q\circ\omega(R_\nabla) \ , \nn\\\SE(1)^2&=&F_\sSE \ .
\label{horizcond}\eea
The superconnection (\ref{curvhomotopy}) thereby defines a continuous
interpolation between the index bundle and the Clifford bundle over $\Sigma$.

The Chern character (\ref{SEChern}) may then be used to construct generalized
Chern-Simons forms via the generating function
\beq
\xi(\hil,\dirac)=\int\limits_0^1dt~\Tr^+\left(\Lambda+\SlashA\right)\,
\exp\frac1{2\pi i}\,\SE(t)^2 \ .
\label{xihildirac}\eeq
The relevant part of (\ref{xihildirac}) insofar as the chiral gauge anomaly is
concerned is the degree 1 component in the BRST ghost field $\Lambda$. To find
it, we use the Duhamel expansion
\beq
\e^{A+B}=\e^A+\sum_{n\geq1}~\int\limits_{\triangle_n}dt_0~\e^{t_0A}\,
\prod_{a=1}^ndt_a~B~\e^{t_aA} \ ,
\label{Duhamel}\eeq
where
\beq
\triangle_n=\left\{\left.(t_0,t_1,\dots,t_n)\in\real^{n+1}~\right
|~t_a\geq0 \ , ~ \mbox{$\sum_a$}\,t_a=1\right\}
\label{simplex}\eeq
is the standard $n$-simplex in $\real^{n+1}$. We set $A=\frac1{2\pi
i}\,(\SlashN^{\rm s}+t\SlashA)^2$ and $B=\frac{t^2-t}{2\pi
i}\,(\omega^2+[\omega,\SlashA])$ in (\ref{Duhamel}). By using (\ref{SEtcurv}),
we then find that the term in (\ref{xihildirac}) which is linear in the BRST
ghost field comes solely from the leading and $n=1$ terms in the expansion
(\ref{Duhamel}). In this way we arrive at the total Chern-Simons form
\bea
&&\xi^{(1)}(\hil,\dirac)=\int\limits_0^1dt~\Tr^+\left[\Lambda\,\exp
\frac1{2\pi i}\,\left(\SlashN^{\rm s}+t\SlashA\right)^2\right.\nn\\&&~~+\left.
\frac{t^2-t}{2\pi i}\,\SlashA\,\int\limits_0^1dt'~\left(\exp
\frac{t'}{2\pi i}\,\left(\SlashN^{\rm s}+t\SlashA\right)^2\right)\,\left[
\Lambda,\SlashA\right]\,\left(\exp\frac{t'-1}{2\pi i}\,
\left(\SlashN^{\rm s}+t\SlashA\right)^2\right)\right] \ . \nn\\&&
\label{xi1hildirac}\eea
Note that when $\alg=R_\nabla=0$, (\ref{xi1hildirac}) coincides with the Cartan
form $\Tr^+\,\Lambda$. In the general case, the form $\xi^{(1)}(\hil,\dirac)$
is a deformation of the Cartan cocycle which has the same group cohomology
class. While this provides a nice characterization of the topological anomaly,
the important aspect is that the inhomogeneous form (\ref{xihildirac}), which
essentially determines the form $\cal Y$ to which the Ramond-Ramond potentials
couple, involves a continuous deformation to the index bundle generated by the
Dirac operator $\dirac$, and is thereby naturally related to K-homology, as we
now explain.

\subsection{Fredholm Modules and Bivariant K-Theory}

To place the discussion of the previous subsection into a precise K-theoretical
framework, we shall use the dual, algebraic characterization of the geometry of
the brane-antibrane worldvolume $\Sigma$ in terms of the algebra ${\sf
A}=C^\infty(\Sigma)$ of smooth complex-valued functions on $\Sigma$. We
represent ${\sf A}$ on the Hilbert space $\hil$ diagonally by pointwise
multiplication of functions. The K-theory group $\K^0(\Sigma)$ may be defined
as the space of equivalence classes of projections $\Pi$ acting on $\hil$. Two
projections $\Pi$ and $\Pi'$ are said to be (algebraically) equivalent if there
is a partial isometry $U$ on $\hil$ with $\Pi=U^\dagger U$ and
$\Pi'=UU^\dagger$. The K-homology group $\K_0(\Sigma)$ is now defined in terms
of Fredholm operators $\fred$ acting on the Hilbert space $\hil$, i.e. those
operators for which there exists another operator $\cal Q$ such that
$\fred\fredadj-\id$, $\fredadj\fred-\id$ and $[\fred,f]^+~~\forall f\in{\sf A}$
are all elements of the elementary algebra $\comp$ of compact operators on
$\hil$. We further assume that the operator $\fred$ is odd with respect to the
$\zed_2$-grading on $\hil$, i.e. $\varepsilon\fred=-\fred\varepsilon$. The pair
$(\hil,\fred)$ is called an even K-cycle and the quadruple $({\sf
A},\hil,\fred,\varepsilon)$ is known as an even Fredholm
module~\cite{Connesbook}. The abelian group $\K_0(\Sigma)$ may be represented
in terms of homotopy classes of K-cycles with respect to direct sum. The
natural pairing between K-theory and K-homology is then provided by the index
map which generalizes (\ref{diracPi}) and (\ref{Indexmap},\ref{Indexmapexpl}),
\bea
\K^0(\Sigma)\times\K_0(\Sigma)&\longrightarrow&\zed\nn\\
\Bigl([\Pi]\,,\,[\fred]\Bigr)&\longmapsto&\ind\,\Pi\,\fred\,\Pi \ .
\label{K0pairing}\eea
Note that any pair $(\hil,\dirac)$ determines a Fredholm module~\cite{Baaj}.
While the Dirac operator $\dirac$ is unbounded, the commutators $[\dirac,f]^+$
are bounded and $f(\id+\dirac^-\dirac^+)^{-1}\in\comp$ for all $f\in{\sf A}$.
Then
\beq
\fred=\frac{\dirac^+}{\sqrt{\id+\dirac^-\dirac^+}}=\dirac^+
\,\int\limits_0^\infty\frac{ds}{\sqrt s}~\frac1{s+\id+\dirac^-\dirac^+}
\label{freddirac}\eeq
is a Fredholm operator. Conversely, any Fredholm module can be obtained in this
way (up to homotopy). The pair $(\hil,\dirac)$ is therefore usually refered to
as a Dirac K-cycle and it is the underlying analytical object which generates
K-homology.

The natural pairing between K-theory and K-homology is actually best understood
through a bivariant form of K-theory known as KK-theory~\cite{KKrev}. The
concept of Fredholm module naturally extends to that of a Kasparov module which
is a quintuple $({\sf A},{\sf B},\hil,\fred,\varepsilon)$ where ${\sf A}$ and
${\sf B}$ are algebras. The generalization which occurs is that while ${\sf A}$
is still represented on $\hil$ by bounded operators, $\hil$ is now a (right)
Hilbert module over ${\sf B}$, i.e. a right ${\sf B}$-module which admits an
inner product with values in ${\sf B}$ and which is complete with respect to
this inner product. The remaining properties are as in the case of Fredholm
modules. The abelian group of homotopy classes of Kasparov modules with respect
to direct sum defines the KK-group $\KK_0({\sf A},{\sf B})$. The functor ${\sf
A}\mapsto\KK_0({\sf A},{\sf B})$ is covariant while ${\sf B}\mapsto\KK_0({\sf
A},{\sf B})$ is contravariant. In particular, when ${\sf B}=\complex$ the group
$\KK_0({\sf A},\complex)$ is by definition just the abelian group of homotopy
classes of Fredholm modules over the algebra ${\sf A}$. Specializing to the
case ${\sf A}=C^\infty(\Sigma)$, we thereby have
\beq
\KK_0\Bigl(C^\infty(\Sigma)\,,\,\complex\Bigr)=\K_0(\Sigma) \ .
\label{KK0homology}\eeq
On the other hand, with ${\sf A}=\complex$ the group $\KK_0(\complex,{\sf B})$
is the abelian group of equivalence classes of (right) projective ${\sf
B}$-modules, and again specializing to ${\sf B}=C^\infty(\Sigma)$ we have by
the Serre-Swan theorem that
\beq
\KK_0\Bigl(\complex\,,\,C^\infty(\Sigma)\Bigr)=\K^0(\Sigma) \ .
\label{KK0theory}\eeq
Therefore, we see that the Dirac K-cycle $(\hil,\dirac)$ interpolates between
the K-homology $[(\hil,\dirac)]\in\KK_0(C^\infty(\Sigma),\complex)$ and
K-theory $\Ind(\hil,\dirac_E)\in\KK_0(\complex,C^\infty(\Sigma))$ groups.

The real advantage of the KK-theory description is that there is typically a
line bundle $\Pi$ which is a projection on
$\orbit_N(C^\infty(\Sigma)\otimes\overline{{\sf A}})$, with $\orbit_N$ the
algebra of $N\times N$ matrices with entries in the given algebra, such that
the index bundle of K-theory can be represented as~\cite{Connesbook,KKrev}
\beq
\Ind(\hil,\dirac)=[\Pi]\otimes_{\overline{{\sf A}}}(\hil,\dirac)\in\KK_0
\Bigl(\complex\,,\,C^\infty(\Sigma)\Bigr)
\label{IndKas}\eeq
in terms of the Kasparov product $\otimes_{\overline{{\sf A}}}$ of $\Pi$ by the
K-homology cycle $(\hil,\dirac)$. The Kasparov product here is a map
\beq
\otimes_{\overline{{\sf
A}}}\,:\,\KK_0\Bigl(\complex\,,\,C^\infty(\Sigma)\otimes
\overline{{\sf A}}\Bigr)\times\KK_0\Bigl(\overline{{\sf A}}\,,\,\complex\Bigr)~
\longrightarrow~\KK_0\Bigl(\complex\,,\,C^\infty(\Sigma)\Bigr) \ ,
\label{Kasprod}\eeq
where $\overline{{\sf A}}$ is the $L^\infty$-norm closure of the algebra ${\sf
A}=C^\infty(\Sigma)$ in the $C^*$-algebra of bounded linear operators on the
separable Hilbert space $\hil$. The result (\ref{IndKas}) emphasizes the fact
that K-homology, through the Dirac K-cycle $(\hil,\dirac)$, is the defining
topological property of D-brane charge.

The Kasparov product is most elegantly described by introducing the notion of a
Cuntz algebra as follows~\cite{Cuntz}. Since $\Sigma$ is assumed to be compact,
${\sf A}=C^\infty(\Sigma)$ is a unital algebra. The Cuntz algebra $Q{\sf A}$ is
defined to be the free product $Q{\sf A}={\sf A}\otimes{\sf A}$ in the category
of unital algebras, i.e. with amalgamation over the identity $\id_{\sf A}$ of
${\sf A}$. Then $Q{\sf A}$ is naturally a super-algebra, and there is a
canonical ``folding'' homomorphism $\varphi:Q{\sf A}\to{\sf A}$ which
identifies the two copies of the algebra ${\sf A}$ inside $Q{\sf A}$. Let
$q{\sf A}=\ker\varphi$. Given a Kasparov module $({\sf A},{\sf
B},\hil,\fred,\varepsilon)$ we can induce a homomorphism $\alpha:q{\sf
A}\to\comp$ by
\bea
\alpha(\eta)&=&P_{\rm GSO}\,\eta \ , \nn\\\alpha^\vee(\eta)&=&
P_{\rm GSO}\,\fred\,\eta\,\fred \ ,
\label{alphadef}\eea
where, for any $f\in{\sf A}$, $f\mapsto f\otimes\id_{\sf A}$ and
$f^\vee\mapsto\id_{\sf A}\otimes f$ are the two canonical monomorphisms ${\sf
A}\hookrightarrow Q{\sf A}$. To characterize the relationship between Cuntz
algebras and KK-theory, we denote by $[q{\sf A},\comp\otimes{\sf B}]$ the
semi-group of homotopy classes of algebra homomorphisms $\alpha:q{\sf
A}\to\comp\otimes{\sf B}$ with respect to the direct sum
$\alpha\oplus\alpha':q{\sf A}\to\orbit_2(\comp\otimes{\sf
B})\cong\comp\otimes{\sf B}$, where the isomorphism is a consequence of Morita
equivalence. Then, for any two algebras ${\sf A}$ and ${\sf B}$, one can show
that~\cite{Cuntz}
\beq
\KK_0({\sf A},{\sf B})=\Bigl[q{\sf A}\,,\,\comp\otimes{\sf B}\Bigr] \ .
\label{KKhomotopy}\eeq
The result (\ref{KKhomotopy}) may now be used to define the Kasparov product
$\KK_0({\sf A},{\sf B})\times\KK_0({\sf B},{\sf C})\to\KK_0({\sf A},{\sf C})$
for any three algebras ${\sf A}$, ${\sf B}$ and $\sf C$.

If $\zeta$ is the canonical generator of
$\K_0(q\complex)\cong\K_0(\complex)\cong\zed$, then to any element
$[\alpha]\in[q\complex,\comp\otimes{\sf A}]$ we can assign the K-theory class
\beq
\K_0(\alpha)[\zeta]\in\K_0\Bigl(\comp\otimes{\sf A}\Bigr)
\cong\K_0({\sf A}) \ ,
\label{K0alpha}\eeq
where we have used the stability of K-theory under Morita equivalence. With
${\sf A}=\complex$ and ${\sf B}=C^\infty(\Sigma)$ we then arrive at the
isomorphism
\beq
\Bigl[q\complex\,,\,\comp\otimes C^\infty(\Sigma)\Bigr]\cong\K^0(\Sigma) \ ,
\label{CuntzK0}\eeq
showing how the Cuntz algebra description (\ref{KKhomotopy}) naturally achieves
the desired K-theory and K-homology interpolation. The natural $\zed_2$-grading
on the Cuntz algebra $Q{\sf A}$ or on $q{\sf A}$ fits in nicely with the fact
that $\Sigma$ is the worldvolume of a brane-antibrane pair, and this natural
association gives rise to the K-homology group $\K_0(\Sigma)$. The crucial
feature here though is that the operator (\ref{superconndirac}) is a Quillen
superconnection and so the corresponding index theorem characterizes the
cohomology class of the topological anomaly, represented most naturally through
the Dirac K-cycle $(\hil,\dirac)$.

The Ramond-Ramond fields $G=dC$ also define elements of K-theory through the
modified Bianchi identity $dG=2\pi\,\ch^+(x)\wedge\sqrt{\widehat{A}(TX)}$ which
leads to the field
\beq
G(x)=2\pi\,\ch^+(x)\wedge\sqrt{\widehat{A}(TX)}
\label{RRKtheory}\eeq
that depends only on the virtual bundle
$x=[(E^+,E^-)]$~\cite{MooreWitten,FreedHop}. The anomalous couplings
$\int_\Sigma{\cal C}\wedge{\cal Y}$ represent a pairing between K-theory and
K-homology, or equivalently a natural pairing on bivariant KK-theory.
Alternatively, we may view it as a pairing between deRham cohomology and
K-homology through the homological Chern character. By using the push-forward
$\phi_*$ induced on homology by the embedding
$\Sigma\stackrel{\phi}{\hookrightarrow}X$, we may lift all of these statements
to find that D-brane charge in Type IIB superstring theory is labelled by the
K-homology group $\K_0(X)=\KK_0(C^\infty(X),\complex)$ of the spacetime
manifold $X$. However, this interpretation does not explain why the geometrical
Dirac genus term appears with a square root in the pairing, as in
(\ref{calYfinal}) or (\ref{RRKtheory}), and the appearence of this factor is
one of the fundamental aspects of the K-theoretic (or otherwise) formulation of
D-brane charges and Ramond-Ramond fields for which a heuristic interpretation
is still lacking.

Analogous conclusions for Type IIA D-branes, for which $\dim\Sigma=p+1$ is odd,
can be reached by removing the grading $\varepsilon$ into positive and negative
chirality spinor fields from the definitions of Fredholm and Kasparov modules
above. The corresponding equivalence classes of odd modules define the higher
K-homology and KK-groups $\K_1(\Sigma)$ and $\KK_1({\sf A},{\sf B})$,
respectively. Again by using push-forward maps $\phi_*$ we may infer that
D-brane charge in Type IIA superstring theory takes values in the K-homology
group $\K_1(X)=\KK_1(C^\infty(X),\complex)$ of spacetime. Note that the
distinction between Type IIA and Type IIB K-groups is far more natural in
K-homology than it is in the K-theory of virtual bundles, because it relies
solely on the dimensionality of the D-brane worldvolume to determine a
chirality grading on the corresponding space of spinor fields. It would be
interesting to find an interpretation of the intermediary bivariant K-theory
groups $\KK_0({\sf A},{\sf B})$ which interpolate between the K-homology and
K-theory groups.

\subsection{Toeplitz Operators and Unstable D-Branes}

Unstable D$p$-branes in Type II superstring theory have been interpreted as
stringy analogs of sphalerons~\cite{Dsphalerons}, i.e. static classical
solutions with a single negative fluctuation mode. They are thereby related to
the non-trivial homotopy of string configuration space, and also intimately to
K-theory. A BPS D$(p-1)$-brane, whose charge is classified by K-theory, gives
rise to a one-parameter family of static configurations whose topology is the
same as that of the stable D$(p-1)$-brane. The extra parameter lives in a
manifold $\cal M$ which replaces the Euclidean time and parametrizes the
D-sphalerons. The unstable D$p$-branes thereby produce a family of virtual
bundles $[(E^+(t),E^-(t))]=x(t)\in\K^0(X)$, $t\in{\cal M}$, so that the
homotopy of string configuration space on the spacetime manifold $X$ is related
to the K-theory group $\K^0(X\times{\cal M})$. The connection to K-theory using
superconnections and index theory, or equivalently Dirac operators, in this
case is most naturally done using the dimensional reduction method of
section~4.1. This interpretation is very much in the same spirit as the
sphaleron picture of unstable branes and it yields an analytical description of
the RR couplings that were derived in section~4.

We work now in Type IIA superstring theory and consider the oxidation
(\ref{hatSigmadef}) of the even-dimensional worldvolume $\Sigma$ of the
unstable D-branes. With the Hilbert space $\hil_{{\bf S}^1}=L^2({\bf S}^1,dy)$,
the Dirac K-cycle of the circle ${\bf S}^1$ parametrized by $y\in[0,1]$ is
$(\hil_\extra,\frac d{dy})$. Its Chern character in homology is the fundamental
class $[\extra]\in H_1({\bf S}^1,\zed)$. We are thereby led to consider the
K-homology cycle of $\hat\Sigma$ which is the graded tensor product
$(\hil',\dirac')=(\hil,\dirac)\otimes(\hil_\extra,\frac d{dy})$, where
$\hil'=\hil\otimes\hil_\extra$ and the Dirac operator of this product is given
by
\beq
\dirac'=(-1)^F\otimes\frac d{dy}+\dirac\otimes\id=\pmatrix{\frac d{dy}&
\dirac^-\cr\dirac^+&-\frac d{dy}\cr} \ .
\label{diractensor}\eeq
The pair $(\hil',\dirac')$ thus defines an odd K-cycle, i.e. it determines an
element of $\K_1(\hat\Sigma)$, and decomposing this group using the K\"unneth
theorem and Poincar\'e duality gives
\beq
\K_1(\Sigma\times\extra)=\K_1(\Sigma)\oplus\K_0(\Sigma) \ ,
\label{Kunneth}\eeq
thereby determining an element of $\K_1(\Sigma)$ via projection onto the first
summand of (\ref{Kunneth}). The details of this projection are equivalent to
the elimination of all higher winding modes of the fields that was done in the
derivation of section~4.1. This follows from the fact that the canonical
projection map $\pi:\hat\Sigma\to\Sigma$ induces an epimorphism
$\pi_*:\K_1(\hat\Sigma)\to\K_1(\Sigma)$ with $\ker\pi_*=\K_0(\Sigma)$, so that
the group $\K_0(\Sigma)$ accounts for the winding modes around the circle
$\extra$ of $\hat\Sigma$.

We can now understand the role of these winding modes more precisely as
follows. The components of the Chern character of $(\hil',\dirac')$ are given
by~\cite{ConnesMosc}
\beq
\ch_n^+(\hil',\dirac')[f_0,f_1,\dots,f_n]=\lambda_n\,
\int\limits_{\hat\Sigma}\widehat{A}\left(
T\hat\Sigma\right)\wedge\Tr\left(f_0~df_1\wedge\cdots\wedge df_n\right) \ ,
\label{chn}\eeq
where $\lambda_n$ are some universal coefficients and
$f_a\in\orbit_N(C^\infty(\Sigma\times\extra))$. The topological anomaly
$\xi_\Lambda$ corresponding to the Cartan-Maurer form
$\Lambda=g^{-1}\,\delta_{\rm BRST}\,g$ will then be determined, as described in
sections 2.4 and 6.1, by a sum over the Chern characters (\ref{chn}) of odd
degree $n=2m+1$ obtained by setting
$f_0=g^{-1},f_1=g,\dots,f_{2m}=g^{-1},f_{2m+1}=g$. Since
$\widehat{A}(T\hat\Sigma)=\widehat{A}(T\Sigma)$, this leads to the
expression~\cite{AS3}
\beq
\xi_\Lambda=\sum_{m\geq0}(-1)^m\,m!\,\lambda_{2m+1}\,
\int\limits_\Sigma\widehat{A}(T\Sigma)\,\oint\limits_\extra
\Tr\left(g^{-1}~dg\wedge dg^{-1}\wedge\cdots\wedge dg\right) \ .
\label{chnanom}\eeq
The integral over the extra dimension $\extra$ in (\ref{chnanom}) is most
elegantly understood through the formalism of Toeplitz operators, as we now
describe.

Let us consider first the case of a single unstable D-brane, i.e. $N=1$. The
Hilbert space $\hil_\extra$ is a module over the algebra ${\sf
A}_\extra=C^\infty(\extra)$, with the action of ${\sf A}_\extra$ on
$\hil_\extra$ given by pointwise multiplication of functions. The Hardy space
$\hil_\extra^+$ is defined to be the $L^2$-norm closure in $\hil_\extra$ of the
linear span of the set of functions $\{\e^{2\pi iny}\}_{n\geq0}$, i.e.
\beq
\hil_\extra^+=\overline{\bigoplus_{n=0}^\infty\complex~\e_{~}^{2\pi iny}} \ .
\label{Hardyspace}\eeq
Let $P_+:\hil_\extra\to\hil_\extra^+$ be the corresponding orthogonal
projection. For any function $f\in{\sf A}_\extra$, we may associate a Toeplitz
operator on $\hil_\extra$ by
\beq
\toe_f=P_+\,f\,P_+ \ ,
\label{Toeplitzdef}\eeq
which, with respect to the orthogonal decomposition
$\hil_\extra=\hil_\extra^+\oplus(\id-P_+)\hil_\extra$, can be expressed as
\beq
\toe_f=\pmatrix{P_+\,f\,P_+&P_+\,f\,(\id-P_+)\cr(\id-P_+)\,f\,P_+&(\id-P_+)
\,f\,(\id-P_+)\cr} \ .
\label{Toeplitzmatrix}\eeq
For any $f\in{\sf A}_\extra$, $\toe_f$ is a bounded operator on $\hil_\extra$.
If $f$ is further an invertible function on $\extra$, then the Toeplitz
operator $\toe_f:\hil_\extra^+\to\hil_\extra^+$ is a Fredholm operator whose
index is given by
\beq
\ind\,\toe_f=\Tr^{~}_{P_+\hil_\extra}\left(\id-\toe_{f^{-1}}\,\toe_f\right)-
\Tr^{~}_{P_+\hil_\extra}\left(\id-\toe_f\,\toe_{f^{-1}}\right) \ .
\label{indextoe}\eeq
It is then a straightforward calculation that establishes the index
theorem~\cite{Toeindex}
\beq
\ind\,\toe_f=\frac1{2\pi i}\,\oint\limits_\extra f^{-1}~df
\label{indextoethm}\eeq
which relates the index of the Toeplitz operator $\toe_f$ to the winding number
of the function $f:\extra\to\extra$.

By considering the Hilbert space $\hil_\extra^+\otimes\complex^N$, this
construction may be easily generalized to matrix-valued functions
$f:\extra\to\orbit_N(\complex)$, and hence to a multi-brane
system~\cite{NCtachyon}. In particular, we may apply an index theorem of the
type (\ref{indextoethm}) to the winding numbers of the tachyon field on
$\hat\Sigma$, regarded as a function $y\mapsto\hat A_y(x,y)$ on $\extra\to
U(N)$. By comparing this result with (\ref{chnanom}), we may now put the index
theoretical calculation of section~4.1 into a proper K-theoretic
interpretation. Namely, the dimensional reduction (obtained by eliminating the
$\extra$-dependence of all fields on $\hat\Sigma$) corresponds to the
incorporation of Toeplitz operators of index zero in the Chern character
mapping of the K-homology cycle over $\hat\Sigma$ which defines the appropriate
index class. However, the crucial property again is that the operator
(\ref{diractensor}) is in fact a Quillen superconnection that acts on smooth
sections of the family of Hilbert spaces $\hil$ over $\extra$. Therefore, the
pertinent index theorem that was used in section~4.1 to yield the unstable
D-brane charge actually computes the overall number of eigenvalues of $\dirac'$
which cross zero in a homotopy between elements in its gauge orbit. The index
theorem thereby again characterizes the cohomology class of the topological
anomaly, or equivalently of the D-brane charge.

It is instructive to see how this structure arises within the formalism of the
brane-antibrane reduction of section~4.2. The Chan-Paton gauge bundle
$E\to\Sigma$ on the unstable Type IIA D-branes is ungraded. However, the
Clifford algebra $\cliff_1^*=\complex\oplus\complex\sigma_1$ has a natural
$\zed_2$-grading, so the product $\hat E=E\otimes\cliff_1^*=\hat E^+\oplus\hat
E^-$ is a superbundle which is identified with the Chan-Paton vector bundle of
the corresponding Type IIB brane-antibrane pairs. Correspondingly, the
endomorphism algebra of $\hat E$ is the superbundle
\beq
{\rm End}\,\hat E\cong{\rm End}(E)\otimes\cliff_1^*
\label{Endcliff}\eeq
which gives rise to superconnections $\hat\alg\in\Omega^-(\Sigma,{\rm
End}\,\hat E)$ of the form $\hat\alg=\id_{2N}\otimes D_A+\sigma_1\otimes T$.
The associated generalized Dirac operators are
\beq
\hat\Sdirac=\pmatrix{\dirac_A&{\cal G}_{\rm s}
\otimes T\cr{\cal G}_{\rm s}\otimes T&\dirac_A\cr} \ .
\label{Sdiraccliff}\eeq
The grading automorphism of the superbundle $\hat E$ is the generator
(\ref{sigmaCliff}) of the Clifford algebra $\cliff_1^*$, and it commutes with
the Dirac operator (\ref{Sdiraccliff}). Thus the pair $(\hil,\Sdirac)$ defines
an {\it odd} K-homology cycle, and hence an element of $\K_1(\Sigma)$. As with
the derivations of section~4, the relationship between the unstable brane
charges and the higher K-homology group of the worldvolume $\Sigma$ comes about
in a much more direct way through the reduction from brane-antibrane pairs. The
dimensional reduction formalism does, however, expose the physical meaning of
the oxidation.

\subsection*{Acknowledgments}

The author would like to thank E. Akhmedov, F. Lizzi, J. Mickelsson, R. Nest,
N. Obers, K. Olsen, B. Schroers, A. Schwarz and S. Willerton for helpful
discussions, and F. Larsen for very helpful comments on the manuscript.
A preliminary version of this paper was presented at the
APCTP-KIAS Winter School and Workshop on ``Strings and D-Branes 2000'' in
Seoul, Korea, February 17--25 2000. Part of this work was carried out during
the PIMS/APCTP/PITP Frontiers of Mathematical Physics Workshop on ``Particles,
Fields and Strings'' at Simon Fraser University, Vancouver, Canada, July 16--27
2001. This work was supported in part by an Advanced Fellowship from the
Particle Physics and Astronomy Research Council (U.K.).

\end{document}